\documentclass[prb,nofootinbib,twocolumn,superscriptaddress]{revtex4-2}

\usepackage{bm,amsmath,mathtools}
\usepackage{graphicx}
\usepackage{xcolor}
  \usepackage{hyperref}
  \hypersetup{
    colorlinks=true,
    hypertexnames=false,
    linktocpage,
    colorlinks=true, 
    urlcolor=magenta!90!black,    
    linkcolor=blue!60!black, 
    citecolor=black!60 
  }

\newcommand{\ket}[1]{\lvert#1\rangle}
\newcommand{\bra}[1]{\langle#1\rvert}

\begin{document}

\author{Jielun Chen}
\affiliation{Department of Physics, California Institute of Technology, Pasadena, CA 91125, USA}

\author{Jiaqing Jiang}
\affiliation{\mbox{CMS, California Institute of Technology, Pasadena,  CA, 91125, USA}}

\author{Dominik Hangleiter}
\affiliation{QuICS, University of Maryland \& NIST, College Park, MD 20742, USA}
\affiliation{JQI, University of Maryland \& NIST, College Park, MD 20742, USA}

\author{Norbert Schuch}
\affiliation{University of Vienna, Faculty of Mathematics, Oskar-Morgenstern-Platz 1, 1090 Wien, Austria}
\affiliation{University of Vienna, Faculty of Physics, Boltzmanngasse 5, 1090 Wien, Austria}

\title{Sign problem in tensor network contraction}

\begin{abstract}
We investigate how the computational difficulty of contracting tensor
networks depends on the sign structure of the tensor entries. Using results 
from computational complexity, we observe that the approximate
contraction of tensor networks with only positive entries has lower
computational complexity as compared to tensor networks with general real
or complex entries.  This raises the question how this transition in
computational complexity manifests itself in the hardness of different
tensor network contraction schemes.  We pursue this question by studying
random tensor networks with varying bias towards positive entries.  First,
we consider contraction via Monte Carlo sampling, and find that the
transition from hard to easy occurs when the tensor entries become
predominantly positive; this can be understood as a tensor network manifestation
of the well-known negative sign problem in Quantum Monte Carlo.  Second,
we analyze the commonly used contraction based on boundary tensor
networks.  The performance of this scheme is governed by the amount of
correlations in contiguous parts of the tensor network (which by analogy
can be thought of as entanglement).  Remarkably, we find that the
transition from hard to easy---that is, from a volume law to a boundary
law scaling of entanglement---occurs already for a slight bias of the
tensor entries towards a positive mean, scaling inversely with the bond
dimension $D$, and thus the problem becomes easy the earlier the larger
$D$ is.  This is in contrast both to expectations and to the behavior
found in Monte Carlo contraction, where the hardness at fixed bias
increases with bond dimension.  To provide insight into this early
breakdown of computational hardness and the accompanying entanglement
transition, we construct an effective classical statistical mechanical
model which predicts a transition at a bias of the tensor entries of
$1/D$, confirming our observations.  We conclude by investigating the
computational difficulty of computing expectation values of tensor network
wavefunctions (Projected Entangled Pair States, PEPS) and find that in
this setting, the complexity of entanglement-based contraction always
remains low.  We explain this by providing a local transformation which
maps PEPS expectation values to a positive-valued tensor network.  This
not only provides insight into the origin of the observed boundary law
entanglement scaling, but also suggests new approaches towards PEPS
contraction based on positive decompositions.  
\end{abstract}

\maketitle

\section{Introduction}

\subsection{Motivation}

Simulating quantum many-body systems is one of the central challenges in
modern physics. Its hardness stems from the exponential size of the
underlying Hilbert space and the presence of  intricate quantum correlations,
which makes it necessary to rely on   suitable approximate methods. Arguably,
the most successful method for a large class of equilibrium problems is Quantum Monte
Carlo (QMC)~\cite{becca:qmc-book}. Yet, especially in systems with strong quantum correlations,
QMC can be plagued by the negative sign problem, or ``sign problem'' for
short. It originates in negative transition matrix elements in the
imaginary time evolution operator $e^{-\epsilon h}$, corresponding to positive
off-diagonal entries in the Hamiltonian terms $h$.

Extensive efforts have gone into finding ways to circumvent or remedy the sign
problem. While in some cases, it can be overcome by a suitable basis
transform, as well as modified QMC algorithms, these approaches are not
successful in the general case~\cite{becca:qmc-book}.  A way out promises the use of other simulation methods
which do not suffer from  a sign problem, such as variational methods. In
particular, tensor network methods have proven instrumental in addressing a range of strongly
correlated quantum many-body problems in recent
years~\cite{banuls:tn-review-23}. For instance,
unlike QMC, tensor network algorithms exhibit no dependence on local basis
choices by construction. 
This raises the question whether methods such as tensor networks can indeed provide a way to circumvent to the sign problem altogether.

In order to address this question, it is first important to understand
what precisely is the difficulty behind the sign problem. Importantly, the
presence or absence of a negative sign problem is not directly equivalent
to computational hardness of the problem, and in particular not to
\textsf{NP}-hardness. Indeed, classical spin  glasses, which do not suffer from the
sign problem, are well-known to be
\textsf{NP}-hard~\cite{barahona:spinglass-np}. In those systems, the hardness
of Monte Carlo algorithms stems from a sampling problem instead, that is, the hardness of setting up a Markov chain which efficiently samples from the correct probability distribution.
Thus, when
discussing how to remedy the sign problem, it is essential to separate
these two sources of hardness, which is challenging since in certain
scenarios, the \textsf{NP}-hardness of a classical problem (i.e., a sampling
problem) can be disguised as a sign problem~\cite{troyer:signproblem-np}.
It is thus central that we are able to single out the intrinsic computational complexity 
of the sign problem.

To this end, we can use insights gained from computational complexity. In
particular, the path sum encountered in QMC is an exponentially big sum, 
where each term (the transition amplitude) is efficiently computable.
Computing such a sum is a defining problem for the class \textsf{\#P} (``Sharp
P''), which amounts to counting the number of solutions to an \textsf{NP}
problem. In the case that the sum only contains positive terms---that is,
in the absence of a negative sign problem---a famous result due to
Stockmeyer~\cite{stockmeyer:approx-counting} (``approximate counting'')
states that this number can be approximated in
$\textsf{BPP}^\textsf{NP}$, that is, by a randomized algorithm with access
to a black-box which can solve \textsf{NP} problems.\footnote{This also has
practical relevance as good heuristic \textsf{NP} solvers exist.} 
In
essence, this takes the complexity of sign-problem free
Hamiltonians back from \textsf{\#P} to the much easier class \textsf{NP}, where also classical
spin glasses lie. 
On the other hand, \emph{general} quantum Hamiltonians are complete
problems for the complexity class \textsf{QMA} (the quantum version of
\textsf{NP}), which is widely believed to be much harder~\cite{kempe:3local-qma,bravyi:stoquastic-ham,gharibian:ham-cplx-book}.
This gap in the computational complexity of general and sign-problem free
Hamiltonians---assuming we believe it exists---will not simply go away
by changing the method used to solve the quantum many-body problem at
hand.

This raises the question as to what happens to the sign problem in tensor
networks---that is, how is the different computational complexity of Hamiltonians with
and without a sign problem reflected in tensor networks? The answer to
this question will depend on how precisely the quantum many-body problem
is mapped to tensor networks. A first possibility is to rewrite the path
sum in QMC as a tensor network in one dimension higher. Then, the sign problem
will exactly correspond to a tensor network which contains non-positive
entries; moreover, it is conceivable that this different sign pattern will
persist when compressing the tensor network in the imaginary time
direction, i.e., truncating the bond dimension, such as done in
simple or full update tensor network algorithms~\cite{jordan:iPEPS,jiang:simple-update}. 

This leads to the question
whether there is a transition in difficulty in evaluating tensor networks
with positive vs.\ general real or complex entries. This is precisely the question which we investigate in this work; 
for a description of the setup, see Fig.~\ref{fig:shiftedgaussian}.

A first insight on the difference in the hardness of contracting tensor
networks with positive entries vs.\ those with arbitrary entries can be
obtained from looking at tensor network contraction as an exponentially
big sum, where each term can be computed efficiently: 
In that case, Stockmeyer's results on approximate counting tells
us immediately that approximating the value of a positive tensor network
is considerably easier (namely, $\textsf{BPP}^\textsf{NP}$).
This perspective also makes it clear that the structure underlying the sign problem in QMC and in tensor network contraction is precisely the same:  
Both are exponentially large sums where each term individually can be efficiently computed, and the hardness of the task is governed precisely by whether the sum is over positive terms or not---the negative sign problem.

However, sampling
is not how tensor networks are contracted in practice: Rather, one
proceeds by contracting the tensor network piece by piece (e.g., column by
column), where at each step, the contracted region is approximated by a
tensor network with lower complexity~\cite{verstraete:2D-dmrg,gray:hyperoptimized-2}. In order to have an efficient
approximation, it is thus necessary to have a low amount of correlations
in the corresponding cuts. The question thus arises whether this way
of contraction is as well subject to a hardness transition when changing between
positive and general entries in the tensor network.  Indeed, such a
transition was recently observed by Gray and Chan when benchmarking
highly optimized tensor contraction routines~\cite{gray:hyperoptimized-2}. Taken together, the above
points strongly motivate a systematic study of a potential hardness
transition in tensor networks as a function of their sign pattern.

\subsection{Results\label{sec:intro:results}}

\begin{figure}
\includegraphics[width=0.95\columnwidth]{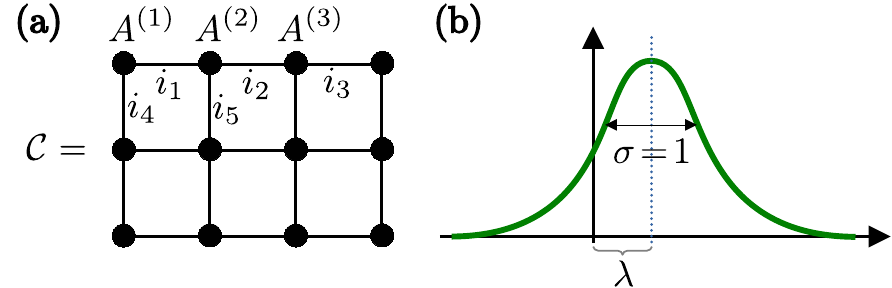}
\caption{\textbf{(a)}
A tensor network is constructed from local tensors $A^{(s)}$. Each leg $i_k$
of the tensor corresponds to an index with \emph{bond dimension} $D$, $i_k=1,\dots,D$. Connected legs are
identified and summed over (``contracted''); after contracting all legs, we
obtain the value $\mathcal C$ of the tensor network, which is a number. We are
interested in approximating $\mathcal C$ up to a given multiplicative error.
\textbf{(b)} 
We consider random tensor networks, where each tensor is chosen Haar-random with
standard deviation $\sigma=1$ (either real or complex), and displaced by a shift
$\lambda$; as $\lambda$ is increased, this gradually transitions to a tensor
network with predominantely positive entries around~$\lambda\approx 1$.}
\label{fig:shiftedgaussian}
\end{figure}

In this paper, we comprehensively study how the difficulty of evaluating
tensor networks depends on the sign pattern of their entries: Is there
a difference in the hardness of contracting tensor networks with only positive
entries as compared to tensor networks with general real or complex entries? How does this
depend on the contraction method chosen? And what determines the
transition between the different regimes?  To address these questions, we study
the average-case difficulty of contracting random tensor networks, that
is, tensor networks with random entries, on a regular square lattice, and with bond dimension $D$, see Fig.~\ref{fig:shiftedgaussian}a.
Here, the entries of each tensor are chosen independently Haar-random
(either real or complex), with standard deviation $1$, and subsequently
shifted into the positive by some amount $\lambda$
(Fig.~\ref{fig:shiftedgaussian}b), allowing us to
smoothly vary from tensor networks with completely random entries (i.e.,
with mean zero) to the regime of positive entries.

First, we consider the contraction of the tensor network using Monte Carlo
sampling, that is, by sampling the exponentially big sum described by the
tensor network, and study when the function which we sum exhibits a
sign problem; the hardness captured by this setup has the same structure
as the sign problem in other Monte Carlo schemes, in particular QMC.  We
observe a transition around $\lambda\approx1$, that is, when the entries
of the network become predominantly positive; this is in line with the
complexity theoretic reasoning we gave above (right before
Section~\ref{sec:intro:results}) based on approximate counting. In
addition, we find that for unbiased (or slightly biased) random tensors with $\lambda\lesssim 1/D$,
the hardness of the problem saturates at the worst-case scaling $D^N$,   
while for larger bias $\lambda$, the severity of the
sign problem decays as a function of $\lambda$ in a way that is independent of the bond dimension. 

We then move on to the hardness of the commonly used entanglement-based contraction---that
is, where the contracted region is approximated by a low-entanglement
state, and the hardness depends on the amount of correlations
(entanglement) in the tensor network. 
We observe that the problem again exhibits a transition from hard to easy,
and thus an entanglement transition from a high-entanglement (volume law)
to a low-entanglement (boundary law) regime. 
In studying this transition,  we however find something extraordinarily surprising:
The transition from hard to easy 
occurs for an ever smaller bias $\lambda$ towards a positive distribution as the bond dimension $D$ is increased.
Specifically, for bond dimension $D$,
a shift $\lambda\approx
1/D$ already induces a transition to a low-entanglement
regime, even though this amounts to a disbalance in the sign pattern of
the order of only $1/D$. We establish the existence of this transition both
through the study of random instances, as well as through the mapping to an effective
statistical mechanics model, which becomes accurate in the large-$D$ limit
and exhibits a transition at $\mu\coloneqq\lambda D=1$ for large $D$. Subsequently, we
demonstrate that this transition indeed has its origin in the shift
towards positive entries, and not in the simultaneously induced decoupling of the network into
local tensor products: To this end, we study 
the two causes separately, and find that they lead to vastly different
effects, regarding both the point and the sharpness of the transition.

As a last point, we return to the use of tensor networks with open
physical indices, in particular Projected Entangled Pair States (PEPS),
as a variational ansatz $\ket\psi$ for quantum many-body wavefunctions.
The central primitive needed to accurately evaluate quantities in
a PEPS $\ket\psi$ is the computation of the normalization $\langle\psi\ket\psi$. We
analyze the difficulty of this problem for random PEPS tensors with complex
entries, and find that both random instances and the mapping to a statmech
model in the large-$D$ limit (for the latter, see also Ref.~\cite{gonzalez:random-tn})
predict that the problem of computing $\langle \psi\ket\psi$ is easy, that
is, the underlying tensor network (without any physical indices) exhibits
a boundary-law entanglement scaling.\footnote{Note that this is different from
the behavior of the physical wavefunction $\ket\psi$, where a volume vs.\ boundary-law
scaling of the state was observed for random states with general vs.\
positive coefficients~\cite{grover:positive-state-entropy}.}
We then give a complexity theoretic underpinning of this result, by
showing that the ``completely positive'' double-layer structure of the
problem can be used to rewrite the contraction as a sum of efficiently
computable positive terms for sufficiently large $D$. This in turn implies that the PEPS contraction can be
approximated much more efficiently than a general sum, using Stockmeyer's
approximate counting (thus taking it from \textsf{\#P} to
$\textsf{BPP}^\textsf{NP}$), as well as through sign-problem free Monte Carlo
sampling. In addition, this provides an alternative explanation for the boundary-law entanglement scaling, by showing that PEPS expectation values are effectively just positive tensor networks, modulo a local transformation.

\section{Setup}

\begin{figure}
\includegraphics[width=\columnwidth]{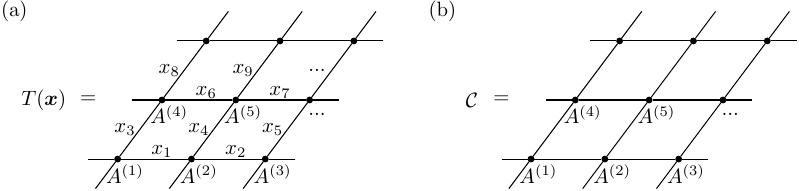}
\caption{
\textbf{(a)} A tensor network (here on a 2D square lattice) is constructed by placing tensors $A^{(s)}_{x_\ell x_rx_ux_d}$ at site $s$. The indices $x_\bullet=1,\dots,D$  are associated to the left, right, up, and down edge adjacent to site $s$, respectively; e.g., the central tensor in panel (a) is $A^{(5)}_{x_6 x_7 x_9 x_4}$. For a given configuration $\bm x=(x_1,x_2,\dots)$ of edges, $T(\bm x)$ denotes the value of the product of all tensors for the given values of $\bm x$. \textbf{(b)} The value of the tensor network $\mathcal C\equiv \mathcal C(\{A^{(s)}\})$ is then obtained by \emph{contracting} the tensor network, i.e., by summing $T(\bm x)$ over all values $\bm x$ of the indices.
}
\label{fig:tn-def}
\end{figure}

Let us start by defining tensor networks, cf.\ Fig.~\ref{fig:tn-def}. A tensor
network, w.l.o.g.\ on a two-dimensional square lattice, is given by a set of four-index
tensors $\{A^{(s)}_{x_\ell x_rx_ux_d}\}_s$, where $A^{(s)}$ corresponds to
site $s=1,\dots,N$, the indices $x_\bullet=1,\dots,D$ (with $D$ called the \emph{bond dimension}) are associated to
edges, and indices of adjacent tensors which are associated to the same edge
$e$ have the same value $x_e$. Let 
\begin{equation}
T(\bm x)\coloneqq\prod_s A^{(s)}_{\bm x}\ ,
\end{equation}
where $\bm x =
(x_e)_e$ contains the value of $x_e$ for all edges in the network, and each $A^{(s)}_{\bm
x}$ only depends on the $x_e$ of the adjacent edges---that is, $T(\bm x)$
is the value taken by the tensor network if each edge is assigned the
value $x_e$.  The value (or contraction) of the tensor network is then given by
\begin{equation}
\mathcal C \equiv \mathcal C(\{A^{(s)}\}) 
\label{eq:contraction}
    \coloneqq \sum_{\bm x} T(\bm x)\ ,
\end{equation}
where the sum runs over all $D^N$ values of $\bm x$. Note that this
construction can be easily adapted to different boundary conditions, by
choosing fewer-index tensors at the boundary when necessary, and similarly
to other graphs.

Tensor networks naturally appear as discretized partition functions
$\mathrm{tr}(e^{-\beta H})$, where the $x_e$
are either configurations at a given point $e$ in space and (imaginary) time;
or as collapsed partition functions, where an edge $e$ corresponds to
a point in space, and $x_e$ holds the configuration at all time slices
simultaneously. For sign problem free Hamiltonians, this naturally
leads to tensor networks with only positive entries (``positive tensor
networks''), while general Hamiltonians will generally carry negative or
complex entries (``non-positive tensor networks''). This raises the
question whether there is a transition in the difficulty of evaluating 
Eq.~\eqref{eq:contraction} 
when transitioning from 
positive to non-positive tensor
networks. Indeed, such a transition has recently been observed by Gray and
Chan in benchmarking tensor contraction codes~\cite{gray:hyperoptimized-2}. 

Before analyzing the reason for this transition, let us specify the
setting we consider. While part of our arguments apply to general tensor
networks, some of them require tensor networks where the tensors
$\{A^{(s)}\}$ are suitably chosen at random. Whenever we discuss concrete
instances, we will draw each tensor from a displaced Haar-random real or complex
distribution, i.e., 
\begin{equation}
\label{eq:A-shifted-orthogonal}
A^{(s)}_{x_\ell x_r x_u x_d} = D^2
U^{(s)}_{(x_\ell x_r x_u x_d)}
+\lambda\ ,
\end{equation}
where $U$ is a Haar-random 
real or complex vector (i.e., the first column of a Haar-random 
orthogonal or unitary matrix $\mathcal U$), respectively.
Here,
the normalization is chosen such that the entries of $D^2U$ have a typical
magnitude of $1$ (indeed, subblocks of $D^2\mathcal U$ converge to a random  real  or  complex
Gaussian ensemble with standard deviation $1$ and mean $0$). 
Adding $\lambda\ge 0$ displaces the distribution, such that the entries of
$A^{(s)}$ become mostly positive (or centered around the positive
axis for complex $U$) for $\lambda \gtrsim 1$.

In the entire discussion, we will always be interested in the difficulty
of approximating $\mathcal C(\{A^{(s)}\})$ up to \emph{multiplicative}
accuracy. This is the natural scenario in particular when we need to take
ratios of tensor network contractions, where good relative accuracies are 
required for the denominator; this setting is natural both in tensor
network simulations and in contexts where sign problems are
relevant in QMC simulations.

\section{Monte Carlo based contraction}
Let us first show that non-positive tensor networks do indeed
exhibit a Monte Carlo sign problem which is not present in positive tensor
networks. To this end, note that Eq.~\eqref{eq:contraction} is an
exponentially big sum, over a discrete configuration space $\bm x$, of a
function $T(\bm x)$ which can be efficiently evaluated for fixed $\bm
x$. Thus, we can use Monte Carlo sampling over $\bm x$ to approximate $\mathcal C$. The
multiplicative error which we obtain for the approximation will depend on
the sign pattern of $T(\bm x)$: If all $T(\bm x)$ point in the same
direction in the complex plane, which does happen around $\lambda\gtrsim
1$, there are no cancellations and we expect rapid convergence.  On the
other hand, for small $\lambda$, where $A^{(s)}_{\bm x}$ and thus $T(\bm x)$ is
distributed around the origin, we expect strong cancellations, and thus a
slow convergence of the relative accuracy of the sample.

The severity of such a Monte Carlo sign problem can be quantified by
comparing the values of the ``fermionic partition function'' $\mathcal C$,
Eq.~\eqref{eq:contraction}, with the ``bosonic'' one, $\mathcal C_b =
\sum_{\bm x}|T(\bm x)|$, where we take the absolute value of $T(\bm x)$~\cite{troyer:signproblem-np,hangleiter:easing-mc-signproblem}. One
expects both quantities to scale exponentially with the volume (i.e., the number 
of lattice sites) $N$, and
thus their ratio should scale with a ``free energy density difference''
$\Delta f$, defined through
\begin{equation}
\label{eq:Delta-f-def}
e^{-\Delta f\, N} \coloneqq \frac{\mathcal C}{\mathcal C_b} = 
\frac{\sum_{\bm x}T(\bm x)}{\sum_{\bm x}|T(\bm x)|}
\end{equation}
for $N\to\infty$.
Since we expect the mean of $K$ samples of $|T(\bm x)|$ to converge as
$1/\sqrt{K}$---but not better, in particular not exponentially fast
either in $K$ or in the system size $N$---and the same 
scaling of fluctuations will 
show up when sampling $T(\bm x)$, a $\Delta f>0$ implies that an exponential
number of sample points $K\sim \exp(2\,\Delta f\,N)$ will be required to get
a good approximation, and thus signals the presence of a sign
problem (which becomes relevant as soon as $N\gtrsim 1/\Delta f$). 

\begin{figure}
\flushleft
\includegraphics[width=\columnwidth]{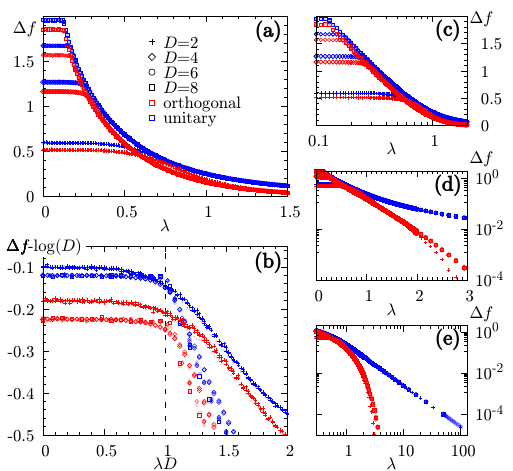}
\caption{
Average free energy difference per site between the value of the tensor
network and its ``bosonic'' version, $\Delta f = \lim_{N\to\infty}
-\tfrac{1}{N}\log(\mathcal C/\mathcal C_b)$, cf.\
Eq.~\eqref{eq:Delta-f-def}, as a function of $\lambda$.
The empty solid-line symbols show data obtained for the distribution
\eqref{eq:A-shifted-orthogonal} for $U$ Haar-orthogonal or -unitary, while
the solid, semi-transparent symbols correspond to the same interpolation
with random Gaussian tensors, see text.  
All panels show the same data, but with different scales. In particular, 
(a) shows the collapse of the data for different $D$ as soon as $\lambda\gtrsim 1/D$; 
(b) the collapse of the lines for all $D$ except except $D=2$, for either given random ensemble, 
to $\Delta f = \log D+\mathrm{const.}$ for $\lambda<1/D$; 
(c) the $\Delta f=\log\lambda + \mathrm{const.}$ scaling for $1/D\lesssim \lambda\lesssim 1$; 
(d) the $\Delta f \lesssim e^{-c\lambda}$ scaling for the orthogonal ensemble for $\lambda\gtrsim 1$; 
(e) the $\Delta f \sim 1/4\lambda^2$ scaling for the unitary ensemble for $\lambda\gtrsim 1$. 
Further discussion on these scalings 
and the implications for the sign problem in Monte Carlo sampling
can be found in the main text. 
}
\label{fig:free-energy}
\end{figure}

Fig.~\ref{fig:free-energy} shows the average $\Delta f$ as a function of
$\lambda$ for different bond dimensions $D$, both for Haar-random
orthogonal and unitary tensors.\footnote{Data has been obtained through
transfer operator calculations, applied to a positive vector, on very long
cylinders of circumference $W$, measuring the ratio of the change in norm
per slice (after discarding the initial head), and taking the average.
Computations for different $W$ show a linear scaling in $W$, implying
stability of the limit $\lim \tfrac1W \log \mathcal C/\mathcal C_b$ used
in obtaining $\Delta f$.} We can identify three different regimes: First,
for $\lambda\le \eta_0/D$ (with $\eta_0\approx 1$), $\Delta f$ is
constant, with a value $\Delta f=\log D + \mathrm{const.}$, as can be seen
from the data collapse for all $D$ except $D=2$ in
Fig.~\ref{fig:free-energy}b (with the constant depending on the choice of
the random ensemble). 
However, once outside this regime, the $\Delta f(\lambda)$
for different $D$ collapse more or less perfectly, with only minor
deviations for $D=2$.  $\Delta f(\lambda)$ then exhibits two different
regimes: For $\eta_0/D\le \lambda \lesssim 1$, we see a behavior $\Delta f
\sim -\log \lambda$ (as expected from the collapse in
Fig.~\ref{fig:free-energy}b and the transition at constant $\lambda D$).
For $\lambda\gg 1$, the scaling behavior depends on the chosen
distribution---for Haar-orthogonal tensors, $\Delta f$ decays
super-exponentially, while for Haar-unitary tensors, $\Delta f\sim
1/4\lambda^2$~(Fig.~\ref{fig:free-energy}d,e).

From Fig.~\ref{fig:free-energy}, we conclude that on the one hand, for
very small $\lambda$, the severity of the sign problem depends on the bond
dimension of the tensor network, getting worse with larger $D$.
Specifically, the $\Delta f=\log D + \mathrm{const.}$ scaling implies that
$e^{\Delta f\,N}\sim D^N$, which is consistent with the expected
worst-case $D^N$ scaling (corresponding to  exact contraction). 
Then, for $1/D\lesssim\lambda\lesssim1$, this transitions into a regime
where there is still a sign problem, which now however is \emph{independent}
of the bond dimension.  Only for $\lambda\gtrsim1$, a transition to a
rapid decay of $\Delta f$ sets in, meaning that the sign problem rapidly
disappears with increasing $\lambda$.  This transition is indeed what
one would have naively expected, given that around $\lambda\gtrsim1$, the
entries of the tensors start to be predominantly positive, with the
probability of negative entries rapidly decreasing as $\lambda$ 
increases.  Concretely, in the orthogonal case, the superexponential decay
of $\Delta f(\lambda)$ implies the absence of a sign problem as long as
$N$ does not grow superexponentially with $\lambda$, while for the unitary
case, the sign problem sets in for $N\sim 4\lambda^2$; we discuss the
origin of this different behavior in the next paragraph. 
To give a specific illustration, for the orthogonal case and $\lambda=2$, a
$10\times 10$ lattice can be contracted using Monte Carlo sampling without
a sign problem, while for $\lambda=3$, one can even go up to a lattice of
about $100\times 100$ sites.
Note that
ultimately, the question whether there is a sign problem depends on the
parameter regime we are interested in: If we keep $\lambda$ fixed and
increase $N$, any setting will eventually exhibit a sign problem (except
for $\lambda\ge D^2$ in the orthogonal case where the probability for
non-positive entries is strictly zero, see below)---yet, the phenomena of
interest are phase transitions in the scaling behavior driven by
collective phenomena, rather than transitions which are due to the
complete absence of negative entries already locally.

Let us now briefly discuss the origin of the different scaling observed in
the $\lambda \gg 1$ regime for Haar-orthogonal vs.\ Haar-unitary tensors,
and its implication on the disappearance of the Monte Carlo sign problem
for $\lambda\gtrsim 1$. To this end, we use that the entries of a
Haar-random matrix follow a Gaussian distribution to high accuracy. For 
Haar-orthogonal tensors, the tensors $A_{\bm x}^{(s)}$,
Eq.~\eqref{eq:A-shifted-orthogonal}, thus follow a Gaussian distribution
with standard deviation $1$ and mean $\lambda$. The probability for a
negative value of $T(\bm x)$ thus originates from the tail of the
Gaussians, which explains their super-exponential decay. On the other hand, for 
a Haar-unitary distribution, the entries $A_{\bm x}^{(s)}=re^{i\phi}$ follow a
\emph{complex} Gaussian distribution with standard deviation $1$ and
mean $\lambda$. The fluctation of the complex phase $\phi$ is thus
$\sigma(\phi)\sim 1/\lambda$ for $\lambda\gg 1$: This implies that when
multiplying $N$ of those numbers, as is the case for the values $T(\bm x)$
of the tensor network for a fixed setting of the indices,  the overall 
phase gets randomized when $\sqrt{N}\sigma(\phi)\sim 1$ and thus,
following Eq.~\eqref{eq:Delta-f-def}, $\Delta f\sim 1/N \sim 1/\lambda^2$.

For comparison, we have also considered tensor networks with real or
complex Gaussian entries (shown in Fig.~\ref{fig:free-energy} with
solid semi-transparent markers), which show the same behavior to very high
accuracy. One difference, however, is the behavior for large $\lambda$ in
the real case, as for Haar-orthogonal tensors, $\Delta f\equiv 0$ as soon
as $\lambda\ge D^2$ (as all entries of a unitary have magnitude $\le 1$),
while for Gaussian tensors, $\Delta f$ will always have some finite
non-zero value due to the tail of the Gaussian, and thus, the decay for
large $\lambda$ must be more rapid; this can indeed be be seen in
Fig.~\ref{fig:free-energy}d,e for the $D=2$ data.

Let us stress once more that all of this is specific to the case where the
tensor network is contracted using Monte Carlo sampling.

\section{Entanglement based contraction\label{sec:ent-based-contraction}}

\begin{figure}
\includegraphics[width=\columnwidth]{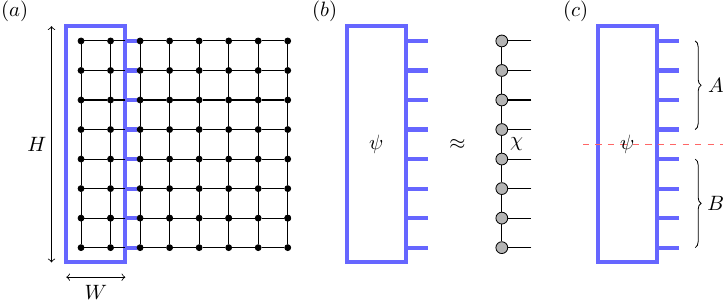}
\caption{
Contraction of a 2D tensor network. One proceeds by blocking columns (blue box
in panel (a)). The resulting block $\psi$ is then approximated by a 1D tensor
network, i.e., an MPS, with bond dimension $\chi$ (panel (b)). In order to keep
the truncation error in all iterations small, it is necessary that the
correlations between the upper and lower part of $\psi$ are bounded, i.e., the
singular value decomposition in the $A$ vs.\ $B$ partition has a small tail
(shown in panel (c) across the central cut, where we expect most correlations.)
This weight can be assessed by whether the entropy across the cut scales with
the width $W$. By choosing $H/2$ sufficiently larger than  $W$, we ensure that
solely $W$ bounds the possible correlations. Concretely, we chose
$H/2=2W$: This value is sufficiently larger than the value
$H/2\approx W$ where the entanglement starts to saturate, and we have checked numerically
that further increasing $H$ does not lead to any change
within error bars. }
\label{fig:boundary-contraction}
\end{figure}

In practice, tensor networks are not contracted using
Monte Carlo sampling.  Rather, in the case of e.g.\ a 2D lattice as in our case, one proceeds by contracting the tensor
network column-wise, where the contracted column tensor $\psi$ is itself again
approximated by a 1D tensor network (``boundary MPS'') with a limited
bond dimension $\chi$. This procedure is illustrated in
Fig.~\ref{fig:boundary-contraction}a,b and further elaborated on 
in the caption. In order to
obtain an accurate boundary MPS approximation, it is necessary that the 
Schmidt coefficients of the boundary tensor $\psi$ 
across all cuts decay rapidly, in a way where the
truncation error does not depend (or only weakly depends) on the width $W$ (where
by analogy, we treat the contracted region as a quantum state $\psi$ which we normalize)---this
ensures an efficient scaling of $\chi$, and thus the computational cost, as the
system size (and thus $W$) increases. 
In particular, we generally expect an efficient scaling in $W$ if the  entanglement across a
cut is constant or grows at most logarithmically in $W$, while a faster growth speaks against such an approximation; 
see also Ref.~\cite{schuch:mps-entropies}.

\subsection{Numerical results}

We have performed numerical studies to see whether the entanglement in the boundary state $\psi$ in the PEPS contraction exhibits a hardness transition when moving towards positive tensor networks. 
To this end, we 
have computed the $2$-R\'enyi entropy 
\begin{equation}
\label{eq:renyi-entropy}
S_2(\psi) =
-\log\frac{\mathrm{tr}_A\big[(\mathrm{tr}_B\,\ket\psi\bra\psi)^2\big]}{
\mathrm{tr}\big[\ket\psi\bra\psi]^2}
\end{equation}
for a contracted
region of width $W$ and a cut in the center of the tensor network, where we expect the entropy
to be largest, see Fig.~\ref{fig:boundary-contraction}a,c. 
Here, $\psi$ labels the tensor for the region, see Fig.~\ref{fig:boundary-contraction}b, 
which we can interpret as an (unnormalized)  bipartite quantum state~$\ket\psi$ between systems~$A$ and $B$.
To ensure that $W$ limits the entanglement between $A$ and $B$, rather than than $H/2$ (the size of the systems $A$ and $B$), we  chose $H/2=2W$; we have checked that the entropy does not grow when $H$ is increased even further.

\begin{figure}
\includegraphics{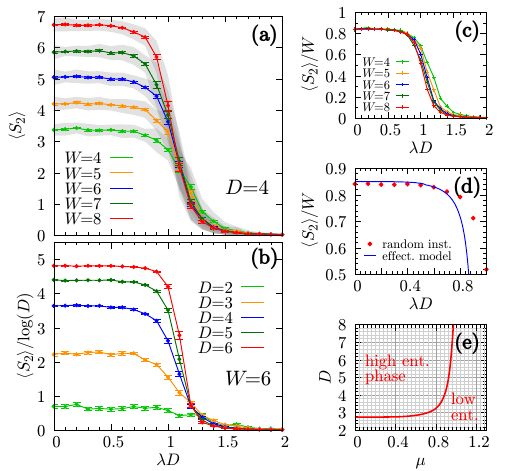}
\caption{
Entanglement (as measured by the $2$-R\'enyi entropy) of the boundary state $\psi$, Fig.~\ref{fig:boundary-contraction}a (with $H=4W$), encountered in entanglement-based contraction of a random tensor network with shifted Haar-orthogonal tensors, Eq.~\eqref{eq:A-shifted-orthogonal}.
\textbf{(a)}~Average entanglement $\langle S_2\rangle$ vs.\ $\lambda D$ for different widths ($D=4$). An increasingly sharp transition from a regime with large entanglement to one with vanishing entanglement is observed around $\lambda D=1$.
The grey stripes indicate the interval in which typical instances (one sigma) lie; the error bars are the standard deviation of the average.
\textbf{(b)} 
Rescaled entanglement $\langle S_2\rangle/\log(D)$ vs.\ $\lambda D$ for different values of $D$ ($W=6$). The plot confirms that the transition happens at constant $\lambda D\approx 1$, independent of the value of $D$ (with the exception of $D=2$, which does not exhibit a transition; see text), and is compatible with 
an expected $S_2\propto \log\,D$ scaling in the highly entangled phase.
\textbf{(c)} Same data as in (a), but with a rescaled $y$ axis $\langle S_2\rangle/W$, which clearly confirms a $S_2\propto W$ scaling in the highly entangled phase; it also shows more clearly how the transition becomes increasingly sharp as $W$ increases.
\textbf{(d)}~Comparison of $\langle S_2\rangle/W$ with the values predicted by the statmech model. Given that the statmech model only reproduces the randomness average correctly in the large-$D$ limit, the data agrees remarkably well (except that the statmech model exhibits a sharper transition).
\textbf{(e)}~Phase diagram as predicted by the statmech model. Note
that the statmech model does not predict a hard phase for $D=2$ even for
$\lambda=0$ (i.e., Haar-distributed tensors with mean $0$), which is in
agreement with the behavior seen in (b), as well as previous
observations~\cite{vasseur:ent-transition-holographic,levy:rand-tn-entropy}.
}
\label{fig:ent-scaling}
\end{figure}

The results are shown in Fig.~\ref{fig:ent-scaling}. Panels a,b,c show data obtained for different values of $W$ and different bond dimensions $D$. The different panels differ by the scaling of the $y$ axis, which we discuss momentarily, while the $x$ axis for all three panels is $\lambda D$.

The plots clearly show a transition in entanglement, and thus a hardness transition in the entanglement-based contraction. What is remarkable, however, is that the transition happens at constant $\lambda D$, roughly around $\lambda D\approx 1$, which becomes evident in Fig.~\ref{fig:ent-scaling}b, which compares data for different $D$ (with the exception of the $D=2$ case, which we will return to later).

The fact that the transition occurs at a displacement $\lambda\approx 1/D$ of the Haar-random (or Gaussian) distribution comes as quite a surprise. 
Both intuitively and from  the analysis in the previous section, we would have expected the transition to occur at a constant value of $\lambda$, independent of $D$, that is, when the entries of the tensors become predominantly positive. Instead, at $\lambda=1/D$, the probability of a positive entry is only roughly $1/2+1/\sqrt{2\pi}D$, barely biased away from a symmetric distribution, and the distribution is becoming more and more symmetric as $D$ is being increased.

Leaving the surprising transition point aside, the entanglement scales as
expected: The collapse in Fig.~\ref{fig:ent-scaling}c (with $S_2/W$ on the
$y$ axis) shows that the entropy in the hard (highly entangled) phase is
proportional to $W$, and Fig.~\ref{fig:ent-scaling}b (with $y$ axis
$S_2/\log D$) is consistent with convergence to an $S_2\propto \log D$
scaling---both are consistent with the maximum possible value $W\log D$
for the entanglement. In the easy (weakly entangled) phase, we observe
(not shown) that the data for different $W$ collapses, and decays roughly
exponentially with $\lambda$, with a slight $D$-dependence of the decay.
Finally, let
us point out that the same transition is also observed in other entropies,
in particular also those with R\'enyi index smaller than one, as well as
for the von Neumann entropy.

Overall, we are left with the puzzling fact that the entanglement transition from hard (volume-law entangled) to easy (boundary-law entangled) happens at a displacement $\lambda$ of the random distribution \eqref{eq:A-shifted-orthogonal} of the tensor which scales as $\lambda\approx 1/D$, and thus a very slight bias of the distribution towards positive entries is already sufficient to reduce the entanglement and make the task of tensor network contraction tractable on average.

\subsection{Randomness average and effective model}
\label{sec:effective-model-orth}

To shed more light on the transition observed in the entanglement, and the location of the transition point at constant $\lambda D$, we study the average behavior of the $2$-R\'enyi entropy, Eq.~\eqref{eq:renyi-entropy},
of the tensor $\psi$, cf.\ Fig.~\ref{fig:boundary-contraction}, where the average is taken over random choices 
of all the tensors $A^{(s)}$ which make up $\psi$---that is, we will consider 
\begin{equation}
\langle S_2\rangle = \int\prod_s\mathrm{d}U^{(s)} S_2(\psi)\ ,
\end{equation}
where the average $\langle\,\cdot\,\rangle$ is over Haar-random $U$ in Eq.~\eqref{eq:A-shifted-orthogonal}. In what follows, we focus on the real-valued Haar-orthogonal distribution; see Sec.~\ref{sec:effective-model-unitary} and Appendix~\ref{appendix:unitary-model} for the Haar-unitary case.

In our analysis, we will concentrate on the limit where~$D$ is large. In that regime, there is a concentration of measure, that is, the fluctuations about the average become small rapidly~\cite{hayden:holographic-duality-from-tn}. This has two important consequences: First, 
the \emph{typical} values, i.e., those which occur with high probability,
will get increasingly closer to the average entropy $\langle S_2\rangle$.
Second, the fact that fluctuations vanish implies that in Eq.~\eqref{eq:renyi-entropy}, we can take the average of the numerator and denominator separately, that is, we can rewrite $\langle S_2\rangle$ as 
\begin{equation}
\label{eq:avg-entropy}
\big\langle S_2(\psi)\big\rangle \approx
-\log\frac{\big\langle
	\mathrm{tr}_A\big[(\mathrm{tr}_B\,\ket\psi\bra\psi)^2\big]
	\big\rangle}{
	\big\langle\mathrm{tr}\big[\ket\psi\bra\psi]^2\big\rangle}\ .
\end{equation}
where the equation becomes exact as $D\to\infty$.

\begin{figure}
\includegraphics[width=\columnwidth]{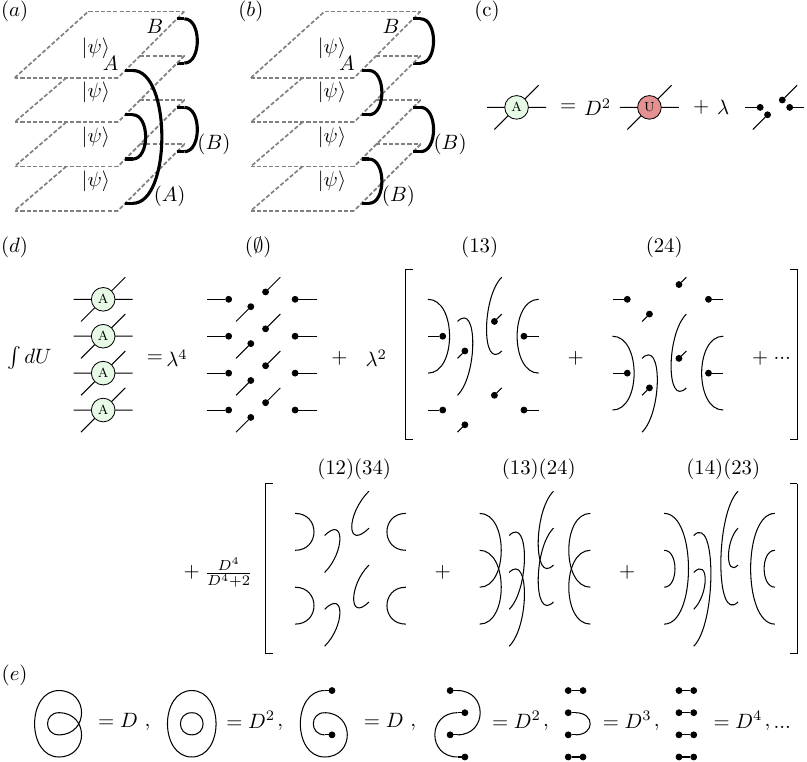}
\caption{Construction of the statistical mechanics model for the average entanglement, Eq.~\eqref{eq:avg-entropy}.
Numerator and denominator are evaluated separately; for either of them, four copies of $\psi$ are required. 
\textbf{(a)} Boundary conditions for the numerator in \eqref{eq:avg-entropy}, amounting to tracing $B$ on twice two copies, squaring, and tracing $A$. 
\textbf{(b)} Normalization in the denominator. 
\textbf{(c)} Tensor used [cf.~\eqref{eq:A-shifted-orthogonal}], which is a sum of a Haar-random tensor and the constant all-$1$ tensor.
\textbf{(d)} The Haar-average over $4$ copies of the tensor $A$ contains terms with none, two, and four Haar-random terms $U$ [cf.~panel (c)], respectively, resulting in all possible ways to pair up the corresponding copies. This gives in total $10$ configurations which can be associated to each site with corresponding weights $\lambda^4$, $\lambda^2$, and $D^4/(D^4+2)$.
\textbf{(e)} Given two configurations at adjacent sites, their contraction gives rise to an overlap pattern of loops and all-$1$ tensors, with some examples shown, where each closed loop or open string evaluates to $D$.
}
\label{fig:statmech-mapping}
\end{figure}

Let us now first consider the Haar-average of the numerator. (We will see later that the analysis of the denominator only differs by boundary conditions.) Computing 
$\mathrm{tr}_A\big[(\mathrm{tr}_B\,\ket\psi\bra\psi)^2\big]$ requires taking 
four copies of $\ket\psi$, $\ket\psi\bra\psi\otimes\ket\psi\bra\psi$ (since for now, we work with real-valued tensors,
we need not worry about conjugates), where suitable boundary conditions---shown in Fig.~\ref{fig:statmech-mapping}a---implement the partial traces and the square.
Since at each site, the average is over the individual Haar-random
$U^{(s)}$, we have to consider 
\begin{equation}
\label{eq:4-copy-expansion}
\begin{aligned}
\int \mathrm{d}U A_{x^1} \otimes & A_{x^2} \otimes A_{x^3} \otimes A_{x^4}
\stackrel{\eqref{eq:A-shifted-orthogonal}}{=}
\lambda^4 \ + \\
&\lambda^2 D^4 \int \mathrm{d}U\, U_{x^1}\otimes U_{x^2} 
+ \dots + \\
&D^8 \int \mathrm{d}U\, 
U_{x^1} \otimes U_{x^2} \otimes U_{x^3} \otimes U_{x^4}\ ,
\end{aligned}
\end{equation}
where now $x^k=(x^k_\ell x^k_rx^k_ux^k_d)$ are the indices of the tensor
in the $k$'th layer, and the middle term includes all
$\genfrac{(}{)}{0pt}{1}{4}{2}=6$ positions for the $U_{x^k}\otimes
U_{x^\ell}$; the odd terms
vanish in the integral. The integrals can be evaluated using Weingarten
calculus~\cite{collins:weingarten1,collins:weingarten2}: Each integral with two $U$'s at $k$ and $k'$ yields
$\delta_{x^kx^{k'}}/D^4$; while the righmost integral evaluates to 
\begin{equation}
    \label{eq:weingarten-4}
\frac{1}{D^4(D^4+2)} \left(\delta_{x^1x^2}\delta_{x^3x^4} + 
    \delta_{x^1x^3}\delta_{x^2x^4} + \delta_{x^1x^4}\delta_{x^2x^3}\right)
\end{equation}
(recall that the dimension of each index $x^k$ is $D^4$).
We can now represent these different terms in the integrals at each site as
configurations $\sigma^{(s)}$ of a classical spin (over which we sum), which is associated to each
site.
This spin $\sigma^{(s)}$ thus takes configurations
\[
1\equiv (\emptyset),\ \{\delta_{ij}\equiv(ij)\},\ \{\delta_{ij}\delta_{kl}\equiv(ij)(jk)\}\
,
\]
that is, it takes $10$ configurations per site (namely $1$, $6$, and
$3$ configurations for the three groups above, respectively). 
These configurations are shown in
Fig.~\ref{fig:statmech-mapping}d. Each of these configurations is assigned a weight from
the prefactor in Eq.~\eqref{eq:4-copy-expansion} (equivalently, in Fig.~\ref{fig:statmech-mapping}d). In addition, contracting the indices
between two adjacent vertices~$s$ and~$s'$ in all layers assigns a weight to each link
which depends on the values of the corresponding $\sigma(s)$ and $\sigma(s')$; specifically,
each open or closed line/loop in the overlap gives a factor $D$, as
illustrated in Fig.~\ref{fig:statmech-mapping}e.

The randomness average thus maps the four copies of 
the tensor network to 
the partition function of a statistical mechanics
model 
\begin{equation}
\sum_{\{\sigma^{(s)}\}} e^{ -\left[\sum_s h(\sigma^{(s)}) + \sum_{\langle
s,s'\rangle} k(\sigma^{(s)},\sigma^{(s')})\right]}\ ,
\label{eq:gen-statmech-Z}
\end{equation}
with one- and two-body terms $h$ and $k$ which depend on~$\lambda$ and
$D$; we return to their specific form in a moment.

In order to compute the Haar-average of 
$\mathrm{tr}_A\big[(\mathrm{tr}_B\,\ket\psi\bra\psi)^2\big]$, 
we still need to
impose suitable boundary conditions, which can be realized by coupling an
additional column of spins in a fixed configuration via $k(\sigma,\sigma')$ to the spins
at the boundary. 
In region $B$ (cf.\ Fig.~\ref{fig:boundary-contraction}c), those spins are in state
$(12)(34)\equiv(B)$, while in region $A$, they are in state
$(23)(14)\equiv (A)$, see Fig.~\ref{fig:statmech-mapping}a.
In the same vein, the normalization can be obtained
by imposing boundary condition $(B)$ everywhere. (Note that $(A)$ and
$(B)$, as well as $(C)\equiv (13)(24)$, are related by permuting copies, so any choice will do.)
In principle, we also need to modify the on-site weights $\exp[-h(\sigma^{(s)})]$ for the spins at the other
boundaries to account for the missing tensor legs, but the precise nature
of those boundaries becomes less important as we increase the system
size, as will become clear when we discuss the mechanism behind the transition.

What is the form of the Hamiltonian terms $h$ and $k$? The derivation, and
its precise form, are given in Appendix~\ref{app:statmech-orth}. What is central for our
derivation is that (after rescaling each tensor by $A^{(s)}\to
A^{(s)}/D$), $e^{-k(\sigma,\sigma')} = \delta_{\sigma,\sigma'} +
O(1/\sqrt{D})$, that is, the system becomes ferromagnetic in the large-$D$
limit. At the same time, $h(\sigma)$ only depends on $\lambda D\eqqcolon\mu$
(together with a weak $D$-dependence of $O(1/D^4)$ which vanishes rapidly
as $D\to\infty)$. This immediately confirms what we observed in
Fig.~\ref{fig:ent-scaling}b, namely that the transition happens around
constant $\lambda D=\mu$---it is the only parameter left in the effective
model. 

So how does changing $\mu$ induce a transition in the system? 
Tuning $\mu$ determines  which configurations are favored by the on-site term $h$ in the Hamiltonian. 
For $\mu<1$, $h$ favors the subspace with the three configurations $(A)$, $(B)$, and $(C)$. 
Since the interaction $h(\sigma,\sigma')$ is ferromagnetic,  this means that the system exhibit symmetry breaking (or, maybe more precisely, long-range order)  in the subspace given by these three configurations. 
That is, up to fluctuations, the spins at all sites want to be in the same state---either $(A)$, $(B)$, or $(C)$. 
We now need to impose the relevant boundary conditions $(A)$ and $(B)$ in the lower and upper half of the system, respectively (Fig.~\ref{fig:statmech-mapping}a). 
This will force the lower half of the system to be in state~$(A)$, and the upper half in state~$(B)$, 
and will thus give rise to a domain wall between the lower and upper half. 
This domain wall will incur an energy penalty proportional to its length, implying an exponential decay of  
the twisted-boundary partition function $\mathrm{tr}_A\big[(\mathrm{tr}_B\,\ket\psi\bra\psi)^2\big]$.
The length scale $\xi$ of the decay depends on the strength of the ferromagnetic interaction $k(\sigma,\sigma')$ (i.e., on $D$, where $\xi\to0$ as $D\to\infty$). 
At the same time, the denominator has the same boundary condition in both regions, and thus does not face such an exponential suppression. 
Thus, the  average R\'enyi entropy (which is minus the logarithm of the above quantity) will increase linearly with the length of the domain wall. 
The length of the shortest (and thus most favorable) domain wall is $W$, and thus, for $\mu<1$ we expect a linear growth of the entropy in $W$.

For $\mu>1$, on the other hand, $h(\sigma)$ uniquely prefers the configuration
$(\emptyset)$, and due to the ferromagnetic coupling, the system will be
(approximately) in the state $(\emptyset)$, and thus disordered. In particular,
this implies that the twisted $(A)$--$(B)$ boundary conditions,
Fig.~\ref{fig:statmech-mapping}a, will give
rise to a \emph{finite} value, independent of $W$. (The precise value will again depend on the strength of the ferromagnetic interactions, that is, on $D$, and vanish as $D\to\infty$.)

We have simulated the phase diagram of the effective statmech model using
infinite MPS (see Appendix~\ref{app:statmech-orth} for details). The phase diagram is shown in Fig.~\ref{fig:ent-scaling}e, where
for large $D$, we can distinguish a symmetry-broken high entanglement phase for
$\mu\lesssim 1$, and a symmetric low entanglement phase for $\mu\gtrsim1$.
For smaller $D$, the high-entanglement phase shrinks, and eventually
disappears for $D\approx2.73$---this suggests that contracting
random tensor networks with $D=2$ is easy on average, and only becomes
hard as $D\ge 3$. While this has to be taken with care, as our analysis
relies on concentration of measure arguments which only become accurate
for large $D$, this is consistent with what we observed in
Fig.~\ref{fig:ent-scaling}b,
as well as with previous findings on
entanglement in Haar-random tensor networks (i.e.\ along the
line  $\mu=0$), where (with a different technique to capture non-integer
bond dimensions) a transition just above $D=2$ had been found~\cite{levy:rand-tn-entropy}.

Let us note that the statmech mapping equally applies to other other lattices or graphs, including higher-dimensional lattices. 
The resulting model only depends on the coordination number $\kappa$ of the lattice, and is obtained from the model above by replacing $D$ by $D^{\kappa/4}$; in particular, in a regular lattice in $\mathcal D$ dimensions, it predicts a transition at $\lambda \approx 1/D^{\mathcal D/2}$.

\subsection{Unitary ensemble\label{sec:effective-model-unitary}}

We have carried out the same analysis---both the study of random instances and the derivation and study of the effective statmech model---also for the complex Haar-random ensemble in Eq.~\eqref{eq:A-shifted-orthogonal}. The findings match those reported in the previous subsections; in particular, the transition is observed at constant $\lambda D$, which can be explained from the effective statmech model obtained from the randomness average, which for large $D$ again exhibits a local term
which only depends on $\lambda D$, together with 
a diverging ferromagnetic coupling. For details, see Appendix~\ref{appendix:unitary-model}.

\subsection{Positive vs.\ factorizing tensor networks}
\label{sec:pos-vs-rank1}

How can we better understand this transition in the entanglement?
One attempt towards an explanation could be to 
interpret \eqref{eq:A-shifted-orthogonal} as an interpolation 
\begin{equation}
\label{eq:S-interpolation}
A^{(s)}_{(x_\ell x_r x_u x_d)}
 = D^2U^{(s)}_{(x_\ell x_r x_u x_d)}+\lambda S_{
(x_\ell x_r x_u x_d)}
\end{equation}
between the Haar-random tensor $D^2U^{(s)}$ and a tensor $S_{x_\ell
x_rx_ux_d}\equiv 1$. We can now write the tensor network as a sum of terms
with either $A$ or $S$ at any site, with a prefactor $\lambda^s$ (where $s$ is the number of times $S$ occurs). Clearly, as $\lambda$ increases, terms
with more $S$ will dominate.  Since $S$ has tensor rank 1,
i.e., is of product form
\begin{equation}
\label{eq:S-rank1}
S_{x_\ell x_rx_ux_d} = v^\ell_{x_\ell} v^r_{x_r} v^u_{x_u} v^d_{x_d}\ ,
\end{equation}
placing an $S$ at a vertex will ``cut apart'' the tensor network at that vertex. For a 
large enough density of $S$ (on the order of one),
we would thus expect a percolation-type transition which would (in typical instances)
decouple the tensor network into efficiently contractible pieces. We
could then e.g.\ try to contract the network by sampling over the distribution
of $A$ vs.\ $S$. At the same time, a realization with many $S$ will
have a small amount of entanglement, and it is not 
implausible that this
property persists even when superimposing all such configurations.

This picture offers a few appealing features, 
and could potentially give rise to
alternative sampling-based contractions. 
It is, however, far from
rigorous -- neither do we know that the sampling does not have a sign
problem, nor is it clear that the fact that most individual realizations of $A$ vs.\ $S$ tensors have low entanglement
is reflected in the entanglement of the overall state. 
At the same time, it suggests interesting connections to e.g.\
measurement-based induced phase
transitions~\cite{skinner:meas-induced-phase-transition,napp:shallow-circuits,chan:unitary-projective-ent-dyn,bao:rand-circuits-meas-transition,jian:meas-induced-criticality-randcirc,ippoliti:meas-only-transition},
where probabilistic (or
otherwise weak) measurements induce an entanglement transition in the
output of a quantum circuit, related to an underlying percolation
transition.

\begin{figure}
\includegraphics{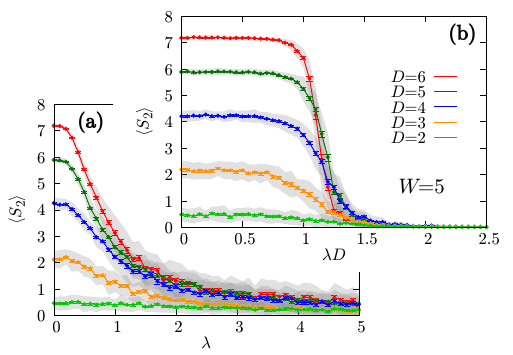}
\caption{%
Average entanglement ($2$-R\'enyi entropy $S_2$) for the interpolation to different tensors, cf.~Eq.~\eqref{eq:S-interpolation}.
\textbf{(a)} Interpolation to a tensor $S$ with rank $1$, but both positive and negative entries:
The transition is now at constant $\lambda$, independent of $D$, consistent with a percolation argument, see text. 
(Specifically, $v^\ell$ and $v^u$ in \eqref{eq:S-rank1} are chosen to have the same number of $+1$ and $-1$ entries, up to rounding, such that the overlaps of two adjacent $S$ yield balanced positive and negative values; note that the sign pattern of the entries of $S$ is only relevant up to rotations on the bonds.)
\textbf{(b)} Interpolation to a randomly chosen site-independent
tensor $S$ (not rank-$1$) with positive entries (chosen uniformly at random between $0$ and $2$). 
The same transition behavior as in Fig.~\ref{fig:ent-scaling} is observed: The transition is much sharper than in panel (a), and occurs at a constant value of $\lambda D$, i.e., $\lambda\propto 1/D$.
}
\label{fig:rank1-interpol}
\end{figure}

If this argument were valid, however, this raises a crucial question: What is
the role of positivity? After all, the argument we just gave should apply to any
rank-$1$ state $S$ [Eq.~\eqref{eq:S-rank1}], even if it is not
positive-valued; and conversely, it does not apply to non-factorizing positive $S$. Fortunately, this can be easily tested numerically, with
the result shown in Fig.~\ref{fig:rank1-interpol} (see caption for
details). 

Specifically, Fig.~\ref{fig:rank1-interpol}a shows the entanglement vs.\ $\lambda$ for an interpolation to a rank-$1$ tensor $S$ with both positive and negative entries.
We observe that---unlike in Fig.~\ref{fig:ent-scaling}b---(\emph{i}) there is no sharp
transition, but rather a relatively smooth decrease in the entanglement
(and, e.g., the tails have much larger entropy),  and 
(\emph{ii}) the cross-over in the entanglement happens around $\lambda\sim1$,
rather than $\lambda=1/D$ (observe the similarity between the curves for
different $D\ge3$). Indeed, this is consistent with the percolation
argument, whose transition depends on the density of $S$, and thus the
value of $\lambda$, independent of $D$
(and indeed, such a percolation transition at fixed finite probability
$p_c$---the percolation threshold---is what is observed also in measurement-induced phase transitions~\cite{skinner:meas-induced-phase-transition,napp:shallow-circuits,chan:unitary-projective-ent-dyn,bao:rand-circuits-meas-transition,jian:meas-induced-criticality-randcirc,ippoliti:meas-only-transition}).
 Still, it is noteworthy that
it is at a value of $\lambda$ on the order of $1$ where the entanglement starts to vanish, 
which is where we also observed a corresponding behavior for
$\Delta f$ (Fig.~\ref{fig:free-energy}), even though in the current case, the transition does 
not seem to depend on  positivity, in contrast to the the transition seen in $\Delta f$.

This shows that for the entanglement transition observed around $\lambda =
1/D$ for the interpolation \eqref{eq:S-rank1}, it is central
that we are interpolating towards positive-valued tensors $S$. 
But does the tensor also need to be of rank~$1$? 
That this should not be a requirement can already be 
argued by rewriting Eq.~\eqref{eq:A-shifted-orthogonal} as an interpolation from
a Haar-random tensor to a mixture of a Haar-random tensor and the identity, by shifting part 
of $U$ into $S$---the resulting model will still have a
transition at a value of $\lambda$ which scales with $1/D$ (depending on the amount shifted).
To substantiate this further, we have also studied an instance where we interpolate towards a fixed random positive tensor $S$ at each site, shown in Fig.~\ref{fig:rank1-interpol}b; just as in Fig.~\ref{fig:ent-scaling}, we observe (\emph{i}) a sharp transition, and (\emph{ii})~at a fixed value of $\lambda D\approx 1$.

The effect of interpolating towards non-positive rank-$1$ tensors $S$ can also be studied once again through a randomness average. To this end, we replace $S$ in Eq.~\eqref{eq:S-rank1} by a Haar-random rank-$1$ tensor (i.e., a product of $4$ Haar-random tensors). The randomness average gives once again rise to a statmech model, which we analyze in Appendix~\ref{app:statmech-rank1}, and which confirms our observations.

\section{PEPS expectation values and completely positive tensor networks\label{sec:peps-contraction}}

In variational PEPS simulations, the tensor networks one needs to contract
generally arise as a normalization or as expectation values of local
observables.  They thus have a very special structure, in that each tensor
$A^{(s)}$ in the 2D tensor network is obtained by contracting the
corresponding PEPS tensor $C^{(s),i}_{y_\ell y_r y_u y_d}$ and its adjoint
through their \emph{physical index} $i=1,\dots,d$ (with $d$ the physical
dimension per site), that is,
\begin{equation}
\label{eq:peps-tensor-CP}
A^{(s)}_{x_\ell x_r x_u x_d} = 
\sum_i C^{(s),i}_{y_\ell y_r y_u y_d}
    \bar{C}^{(s),i}_{z_\ell z_r z_u z_d}
\end{equation}
(see Fig.~\ref{fig:cp-contraction}a),
where the double-layer index $x_\bullet = (y_\bullet,z_\bullet)$,
$\bullet=\ell,r,u,d$, has
dimension $D^2$ for a PEPS with bond dimension $y_\bullet=1,\dots,D$.
The tensors $A^{(s)}$ of the resulting 2D tensor network thus do
not in general have positive entries (unless 
they are 
obtained from PEPS with positive entries).
They do, however, come with \emph{another} type of positivity: 
When
interpreted as a map from ket to bra indices, they amount to positive
semidefinite operators, i.e., they can be written in the form
\eqref{eq:peps-tensor-CP} (i.e., ``$C^\dagger C$'' acting from ket to bra) 
for some $C^{(s)}$.  
This is also true for computing expectation values of of physical
observables, since any observable can be expressed as the difference of two
positive semidefinite operators (and even for string operators, those
positive semidefinite operators can be decomposed in terms of a
double-layer tensor network).
This raises the question whether this positivity structure 
can still ease
the computational difficulty of the contraction.  
In the following, we will  address this
question through the various tools introduced in the preceding sections.

\subsection{Randomness average}

First, we can map the
problem to a statmech model, by taking Haar-random tensors $C^{(s)}$ and
taking the randomness average of four copies of the resulting 2D tensor
network defined by $A^{(s)}$ [Eq.~\eqref{eq:peps-tensor-CP}], which amounts to 
$\big\langle(C^{(s)}\otimes \bar{C}^{(s)})^{\otimes 4}\big\rangle$. 
In the given situation, the restriction to
real-valued $C^{(s)}$ seems less motivated, and we thus focus on Haar-random
unitary tensors $C^{(s)}$. The resulting Haar-average is a sum
over all permutations $\pi\in S_4$ of four copies, where each ket degree of
freedom in copy $k$ is paired up (through a Kronecker delta) 
with the corresponding bra degree of freedom in copy
$\pi(k)$. The state space per site of the resulting statmech model is thus
labelled by permutations $\pi\in S_4$. 

Now consider a site in state $\pi$. The contracted physical degrees of freedom of the
four copies thus form loops corresponding to the cycles of $\pi$; we
denote the number of cycles by $c(\pi$).  Each closed loop contributes a
factor $d$ to the Gibbs weight, and thus, the resulting Hamiltonian is 
\begin{equation}
h(\pi)=-\log(d)\, c(\pi)
\end{equation}
(corresponding to a Gibbs weight $e^{-h(\pi)}=d^{c(\pi)}$). On the other
hand, contracting two adjacent sites in states $\pi_1$ and $\pi_2$ gives
rise to a pattern of closed loops described by the difference
$\pi_1\pi_2^{-1}$ of the two permutations, where now each closed loop
contributes a factor $D$; the resulting interaction term in the
Hamiltonian is thus given by 
\begin{equation}
k(\pi_1,\pi_2) = -\log(D)\,c(\pi_1\pi_2^{-1})\ .
\label{eq:S4-interaction}
\end{equation}
Together, these two terms describe the effective statmech model, cf.\
Eq.~\eqref{eq:gen-statmech-Z}. 
We note that a closely related model for the Gaussian ensemble was derived independently in Ref.~\cite{gonzalez:random-tn}.

Let us now analyze this model. Clearly, $c(\pi)=4$ for the identity
permutation $\pi=\mathrm{id}$, and $c(\pi)\le 3$ otherwise.  In the large
$D$ regime, the interaction term \eqref{eq:S4-interaction} thus describes
a ferromaget with a $24$-fold degenerate ground space. 
For physical dimension $d=1$, the local term $h$ vanishes, and the model
therefore exhibits symmetry breaking between the boundary configurations
relevant for computing the R\'enyi entropy
(cf.~Fig.~\ref{fig:statmech-mapping}a), which gives rise to a domain wall
and as a consequence a linear growth of the entanglement entropy. This is
indeed what we should have expected, since for $d=1$, what we study are
simply two copies of the same random tensor network with complex entries,
for which had already established that it displays a linear growth of 
entanglement. 

What happens when we set $d>1$?
Then, 
the on-site term $h(\pi)=-\log(d)c(\pi)$ will favor the state $\pi=\mathrm{id}$ at each
site. This immediately lifts the degeneracy of the ground space, and the
system will be in the state $\pi=\mathrm{id}$ at all sites, leading to a
breakdown of long-range order. (This is due to the fact that the model was
at a first-order transition.) From the analysis of the statmech model we
thus expect that, at least for sufficiently large $D$, the problem of
contracting PEPS (that is, contracting the double-layer tensor network)
 will not exhibit a linear growth of entanglement, and should thus
scale efficiently again.  

\subsection{Numerical simulation}

\begin{figure}
\includegraphics[width=\columnwidth]{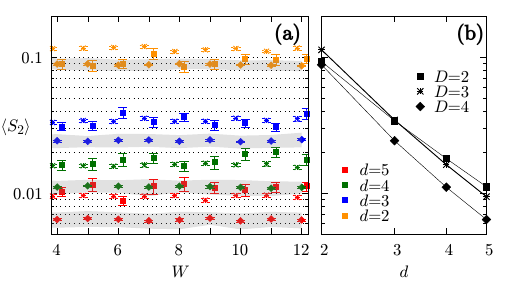}
\caption{
Entanglement scaling for the contraction of PEPS expectation values. Both panels show the average $2$-R\'enyi entanglement entropy $\langle S_2\rangle$ of the boundary state $\psi$, cf.\ Fig.~\ref{fig:boundary-contraction} (for $H=4W$), averaged over $50$ random instances. \textbf{(a)} Dependence of $S_2$ on $W$, for different bond dimensions $D$ and physical dimensions $d$; we find that the entanglement does not change with $W$. 
The grey stripes show the interval for typical (one sigma) values of the entropy
for $D=4$.
\textbf{(b)} Dependence of $S_2$ on the physical dimension $d$ (averaged over $W=8,\dots,12$). We find an algebraic decay with exponent $\alpha\approx 3$; the dependence on $D$ is comparatively weak.
}
\label{fig:PEPS_entropy}
\end{figure}

This behavior of the entropy is confirmed by numerical simulations
of random instances. As shown in Fig.~\ref{fig:PEPS_entropy}a, 
the entanglement of $\psi$ (Fig.~\ref{fig:boundary-contraction}) 
is essentially independent of $W$. Fig.~\ref{fig:PEPS_entropy}b shows the
dependence of the entropy (averaged over widths $W=8,\dots,12$) on $d$, where we
observe an algebraic decay $S_2\propto d^\alpha$ (with $\alpha\approx 3$) of the
entropy with the dimension of the physical system, while the dependence on the
bond dimension $D$ is comparatively minor. The origin of the $d$-dependence will
become more clear in the following subsection.

\subsection{Expressing PEPS expectation values as a sum of positive terms\label{sec:pepscontraction-tnr-proof}}

In the case of positive-valued tensor networks, we were able to argue that
they should be more easily contractible by expressing the contracted value
$\mathcal C$ as a sum over all classical configurations of the contracted
bond variables, Eq.~\eqref{eq:contraction}. Since each of the terms in the
sum was positive, $\mathcal C$ could be approximated using
Monte Carlo sampling without a sign problem. On a formal complexity theoretic level, applying approximate counting allowed to show that the complexity of approximating $\mathcal C$ had a significantly reduced computational complexity. For the current case of
contracting PEPS expectation values, the same approach will not work:
Sampling over classical configurations of the bonds gives non-positive
terms, and sampling over a basis of positive operators $\rho_i$ (in the ket-bra
partition) instead is not possible since there is no resolution of the
identity in terms of only positive operators (the map $\sum
\rho_i\mathrm{tr}[\sigma_i\,\cdot\,]$ with $\sigma_i,\rho_i\ge0$ is entanglement breaking).

\begin{figure}
\includegraphics[width=8cm]{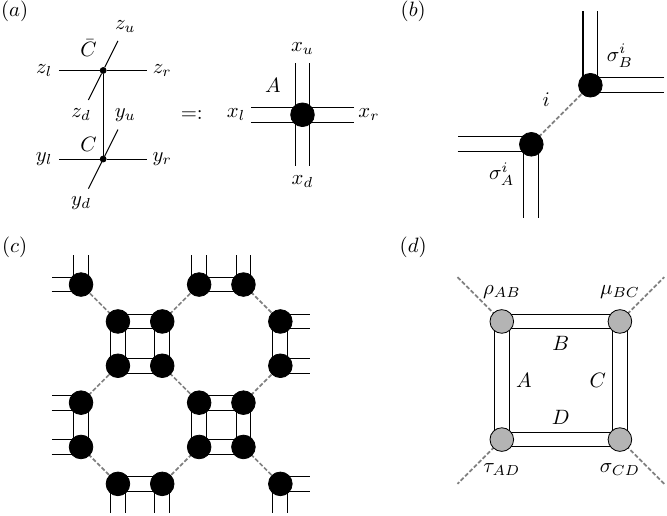}
\caption{
Decomposition of a PEPS expectation value as a sum of efficiently computable
positive terms. \textbf{(a)} The two-layer PEPS tensor can be interpreted as
tracing out the physical index of a five-partite state $C$, resulting in a mixed
state $A$ acting from the ket (bottom) to the bra (top) layer. The right hand
side of the equation shows a top view of the double-layer tensor $A$, where the
double legs indicate the presence of a ket and bra layer; this notation is  also
used in the other panels. \textbf{(b)} For large enough dimensions and
Haar-random tensor $C$, $A$ is close to the maximally mixed state, and thus has
a separable decomposition (dashed line). \textbf{(c,d)}~The separable decompositions are
chosen in alternating partitions, such as to form small loops. The value of such
a loop can be efficiently computed, and is always positive (as it is the overlap
$\mathrm{tr}[(\rho_{AB}\otimes\sigma_{CD})(\tau_{AD}\otimes \mu_{BC})]$ of two
positive operators). Instead of loops, other efficiently contractible structures
such as strings or trees also work fine. Throughout the argument, density
operators act between ket and bra indices.
}
\label{fig:cp-contraction}
\end{figure}

Yet, as we show in the following, it is possible to rewrite the
contraction $\mathcal C$ of a PEPS expectation value as a sum of
positive terms, which can in turn be evaluated by sign-problem-free Monte
Carlo sampling, and allows for the application of approximate counting to reduce the complexity from \#\textsf{P} to \textsf{BPP}$^\textsf{NP}$; the argument works with probability approaching one for sufficiently large bond and
physical dimension. To this
end, observe that the single-site double-layer tensor $A^{(s)}$ can be
interpreted as a density operator $\rho$ between the ket and the bra layer of the tensor network, which is obtained from a Haar-random
state $C^{(s)}$ by tracing out the physical index (i.e., contracting it with its adjoint, Fig.~\ref{fig:cp-contraction}a). If we further group the
tensor legs into two groups of two (e.g.\ left+down vs.\ up+right, Fig.~\ref{fig:cp-contraction}b), we can think of $\rho$ as a bipartite mixed state
$\rho_{AB}$ defined on a Hilbert space of dimension $D^2\times D^2$, which is obtained
from a Haar-random pure state of dimension $d\times D^2\times D^2$ by tracing out the $d$-dimensional system. We can now combine two facts: First,
if we trace a large enough part of a high-dimensional Haar-random state,
the remaining state $\rho_{AB}$ is close to the identity~\cite{hayden:generic-ent}; and second, if a
state is sufficiently close to the identity, it is separable~\cite{braunstein:separable}, i.e.,
\begin{equation}
\label{eq:separable-dec}
\rho_{AB}=\sum p_i \sigma_{A}^i\otimes \sigma_B^i\ ,
\end{equation}
with
$\sigma_\bullet^i\ge 0$ and $p_i>0$; see Fig.~\ref{fig:cp-contraction}b for an
illustration.
More specifically, this
transition occurs when $d\sim D^6$~\cite{aubrun:entanglement-threshold},
with probability
approaching $1$ as $D\to\infty$. Moreover, the decomposition
\eqref{eq:separable-dec} can be constructed efficiently in the dimension of the
space~\cite{braunstein:separable}.\footnote{E.g., by first expanding
$\rho_{AB}$ in a hermitian basis for $A$ and $B$ which
includes the identity, and then pairing each term with part of the identity to make
it positive.}

We can now place such a separable decomposition of the tensor $A^{(s)}$ at each vertex, where we
choose the bipartition such that the grouped indices 
form squares, cf.\ Fig.~\ref{fig:cp-contraction}c. 
Let us now choose a specific value $i_k$ in the
sum for the separable decomposition \eqref{eq:separable-dec} at each site.
Then, 
\begin{equation}
\mathcal C = \sum_{i_1,\dots,i_N} T(i_1,\dots,i_N)\ ,
\end{equation}
where the value $T(i_1,\dots,i_N)$ is obtained by multiplying the overlaps
of the separable decompositions, which can be calculated individually 
across every other
plaquette. For each plaquette, this value is of the form
$\mathrm{tr}[(\rho_{AB}\otimes\sigma_{CD})(\tau_{AD}\otimes \mu_{BC})]$
(see Fig.~\ref{fig:cp-contraction}d),\footnote{The operators $\rho_{AB}$, $\sigma_{CD}$, $\tau_{AD}$, $\mu_{BC}$ are positive semidefinite operators acting from the ket to the bra layer of the pair of legs indicated in the subscript.}
which is an overlap of two positive semidefinite operators and thus
itself non-negative. We thus find that the contraction $\mathcal
C$ of the tensor network can be expressed as an exponentially big sum over
efficiently computable positive terms. Therefore, it can be efficiently
approximated using Monte Carlo sampling without a sign problem, and approximate counting implies that it can be approximated with a much lower computational complexity.

The fact that for $d\gg D^4\gg1$ a Haar-random
tensor becomes close to maximally entangled between virtual and physical indices is
equivalent to saying that the PEPS tensor (as a map from virtual to physical
indices) gets close to an isometry. This means that the PEPS approaches a state
which simply consists of maximally entangled bonds  between adjacent sites, with
no non-trivial map acting on them (that is, a Projected Entangled Pair State
without a projection). One can then immediately see why such a state does not
exhibit any entanglement between different regions $A$ and $B$ at the boundary,
just as observed in the preceding section.

Let us emphasize that we do not expect to be able to obtain a positive sum
decomposition for the expectation value of every PEPS, as there is an intrinsic
gap in their worst-case computational complexity: Computing the normalization of
a PEPS is a \textsf{GapP}-hard problem (that is, an exponentially big sum with
positive and negative weights), or equivalently \textsf{PostBQP} (quantum
computing with postselection)~\cite{schuch:cplx-of-PEPS,aaronson:postsel}, which
is not expected to be more efficiently approximable, in contrast to
exponentially big positive sums (i.e.,
\textsf{\#P})~\cite{stockmeyer:approx-counting}. It remains however an
interesting problem for future research to investigate if the above
decomposition of a random tensor network into a positive sum can be extended to
random tensor networks with any bond dimension, for instance by blocking
regions.  

The fact that 
computing the normalization of a random PEPS can be
rewritten as a positive sum---in fact, the contraction of a positive
tensor network---for sufficiently large dimensions has an interesting consequence: It implies that
under this rewriting, the contraction of the PEPS reduces to the
contraction of a positive tensor network, which---as we have found in Section~\ref{sec:ent-based-contraction}---does
indeed have a favorable entanglement scaling. The above mapping thus 
provides an additional argument as to why the entanglement-based
contraction of PEPS scales nicely despite the presence of non-positive
entries: PEPS expectation values are really just positive tensor networks in
disguise. 

Note that the transformation in
Fig.~\ref{fig:cp-contraction} can be interpreted as one step of a kind of tensor network renormalization (TNR)
algorithm~\cite{evenbly:tnr}; it is then an interesting question to further investigate how TNR can be used to remove negative signs in tensor networks, and whether optimizing for positivity provides a viable route towards a stable and efficient application of TNR to the computation of PEPS expectation values (which usually suffers from instabilities due to the dependency on gauge choices~\cite{banuls:private}).

\section{Summary and Discussion}

\subsection{Summary}

In this work, we have considered the hardness of contracting tensor
networks, and investigated how it depends on the choice of tensor entries,
namely positive entries vs.\ general real or complex entries.  We used
complexity theoretic considerations to argue that there must be a
transition in the computational complexity of the problem as the sign
pattern changes. We then investigated how this ``sign problem'' would show
up in different numerical contraction methods, by considering random
tensor networks with entries with mean $\lambda$ and standard deviation
$\sigma=1$. On the one hand, we considered contraction by Monte Carlo
sampling. There, we found a gradual transition in the hardness of the
problem around $\lambda=1$, which is where the tensor entries become
predominantly positive. This is in agreement with expectations, and can be
seen as a tensor network version of the well-known Quantum Monte Carlo sign
problem. On the other hand, we investigated the commonly used contraction
method based on boundary MPS. We found that it exhibits a much earlier
transition: The entanglement in the boundary state undergoes a transition
from volume law to boundary law entanglement scaling already at a mean of
$\lambda\approx 1/D$, much earlier than the transition in the Monte Carlo
sign problem, and, most surprisingly, earlier for larger $D$. We observed this
behavior in the study of random instances, and were able to confirm
it through the construction of an effective classical statmech model. We were
also able to show that the entanglement transition is governed solely by the
shift towards positive entries, and occurs also when interpolating to a
correlated random tensor with positive entries.  Finally, we investigated the
contraction of PEPS expectation values, where we found a boundary law
scaling of the entanglement independent of the physical and bond
dimension;  we subsequently showed that in a certain regime, such a
contraction could indeed be rewritten as a positive-valued tensor network,
proving its reduced computational complexity, and at the same time 
providing a direct intuition for the observed entanglement scaling.

\subsection{Context}

Entanglement transitions have been studied in recent years in a variety of
contexts. This in particular involves the questions of dynamics and
equilibration, such as the MBL transition~\cite{nandkishore:mbl-review} or
entanglement transitions in monitored quantum dynamics,
measurement-induced transitions, or other transitions obtained by adding
certain non-unitary gates to unitary quantum circuits
\cite{skinner:meas-induced-phase-transition,napp:shallow-circuits,chan:unitary-projective-ent-dyn,bao:rand-circuits-meas-transition,jian:meas-induced-criticality-randcirc,ippoliti:meas-only-transition}.
Since quantum circuits, including measurements, can be expressed as tensor
networks, these systems therefore also provide examples of entanglement
transitions in tensor networks. Similarly, entanglement transitions in
random tensor networks as a function of the bond dimension $D$ have been
investigated previously for random entries with mean
$0$~\cite{vasseur:ent-transition-holographic,levy:rand-tn-entropy} (i.e.,
the subfamily with $\lambda=0$ of the class we investigated), mimicking
non-integer $D$ by a suitable random distribution of bond dimensions;
there, an entanglement transition between $D=2$ and $D=3$ has been
observed, consistent with our findings. 

This illustrates that there is a number of different parameters which can
drive entanglement transitions in tensor networks. On the one hand, our results add another
axis to it, namely, the positivity of the tensor entries. Yet, there is a
profound difference in the nature of the sign problem transition which we
observe: The root of this transition is in the fundamentally different
computational complexity of the underlying computational problem, and the
observed entanglement transition is only one possible way in which this
complexity transition manifests itself. In fact, in a forthcoming
work~\cite{jiang:positive-tn}, we will show that the same transition also
appears in an entirely different contraction algorithm which is inspired
by Barvinok's method for approximating the
permanent~\cite{barvinok:permanent-complex,eldar:permanent}, which
makes no reference to entanglement whatsoever, and which can be rigorously
proven to work. In this context, a key question which remains to be
answered is to determine the true point of the complexity transition, that
is, to identify a regime for $\lambda$ where contraction is provably hard.

A related question connects to the QMC sign problem, where the question we posed
initially originated from. Our findings do not conclusively answer what
happens to the QMC sign problem in tensor networks, i.e., where the
intrinsic computational complexity of simulating Hamiltonians with vs.\
without a sign problem (i.e., \textsf{QMA}-hard vs.\ \textsf{stoqMA})
shows up in tensor network algorithms. This has several reasons: First, we
have focused on random tensor networks, and the tensor networks which
appear in numerical simulations are generally not random. (For instance,
it is easy to come up with tensor networks with only positive entries with
a volume law, by setting indices to be equal such that they form long
strings, sometimes termed ``rainbow state''.) Second, we have only
considered the contraction of tensor networks, as it e.g.\ arises from
iteratively truncating the imaginary time evolution in a simple or full
update algorithm, but which does not take into account the computational
effort required to find the optimum in variational tensor network
algorithms. Finally, as we have observed in
Section~\ref{sec:peps-contraction}, computing expectation values at least
for random PEPS has a favorable entanglement scaling again, leaving open
the possibility that in variational PEPS simulations, the difficulty of
solving a problem with an intrinsic sign problem is entirely in the
minimization itself. Understanding what happens to the complexity of a
Hamiltonian with an intrinsic QMC sign problem when tackling it with tensor network
algorithms thus remains an important open problem. However, one should
keep in mind that this question is likely a very hard one, and will depend
sensitively on the chosen algorithm, since already in one dimension, it
can on the one hand be shown that e.g.\ certain variants of the
variational DMRG algorithm can get  stuck in local
minima~\cite{eisert:DMRG-NP,schuch:mps-gap-np}, while on the other hand,
provably converging algorithms can be
devised~\cite{landau:1d-efficient-algo}.

\subsection{Practical implications}

Finally, let us discuss how the results of this work, which focused on
the study of random tensor networks, can be relevant for practical
settings, in particular for the simulation of physical many-body problems
based on tensor network methods.

A first point to note is that random tensor networks with positive
entries, as considered in this work, naturally appear as partition
functions of random classical statistical mechanics models, which are
widely studied in the context of disordered many-body systems. Thus, the
results of the analysis of random tensor networks which we obtained in this work
directly carry over to random classical statistical mechanics models.

Next, let us turn towards the possible implications of our results for
numerical simulation algorithms for quantum many-body systems based on
tensor networks, in particular variational PEPS simulations. A key
insight of our work is that positive tensor networks---and, in fact,
already tensor networks which are biased towards positive entries---possess 
a much lower amount of entanglement and can therefore be more
efficiently contracted using commonly used contraction algorithms. This suggests that positivity of tensor entries
can serve as a guide in choosing the most suitable gauge when contracting
tensor networks in PEPS simulations: Gauge choices can have a significant
effect on the efficiency of contracting a tensor network, and at the same
time, choosing the right gauge is a challenging problem in many cases.

\begin{figure}
\includegraphics{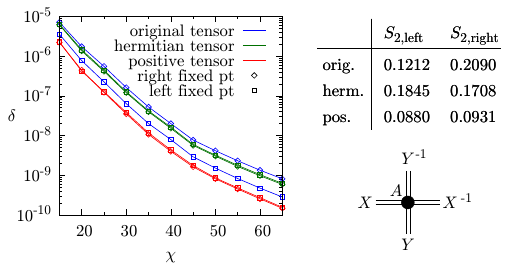}
\caption{\label{fig:gaugefixing}
Contraction of a translational invariant infinite PEPS obtained by
optimizing a highly frustrated topological spin liquid (square lattice
$J_1$-$J_2$ model at $J_2/J_1=0.5$) which exhibits a strong sign problem.
The plot on the left shows the truncation error $\delta$ (sum of discarded
Schmidt coefficients) for the left and right transfer matrix fixed point
as a function of the truncation dimension $\chi$ (cf.\
Fig.~\ref{fig:boundary-contraction}b), while the
table on the top right shows the $2$-Renyi entanglement entropies $S_2$ for a
bipartition of those fixed points. We compare data obtained for the original iPEPS
tensor returned by a variational minimization (blue) with data obtained
for tensors where we have optimized over gauge choices $X$ and $Y$ as shown
on the bottom right. On the one hand, we have optimized $X$ and $Y$ such as to
maximize the hermiticity of the transfer matrix (green), and on the other hand, we have
additionally optimized $X$ and $Y$ for positivity of the tensor entries
(red).
We see that both optimizations lead to a situation where the truncation
error and entanglement entropy for the two fixed points become symmetric.
However, while optimizing for hermiticity leads to an increased truncation
error and entanglement entropy for one of the fixed points (such that the
optimized, symmetric values are in between the original values), optimizing for
positivity allows to decrease the truncation error for both fixed points
simultaneously and thus gives rise to a more efficient contraction of the
tensor network.  
}
\end{figure}

One possibility how positivity of tensor entries can guide such a gauge
choice is as follows (for clarity, we focus on the practically most relevant
case of a translational invariant infinite tensor network). We start from the
two-layer tensor network obtained when computing expectation values, which
is made up from individual two-layer ket-bra tensors $A$ just as the one 
shown in Fig.~\ref{fig:cp-contraction}a. The goal is to contract this
tensor network on an infinite system using infinite boundary MPS (iMPS)
(i.e., the translational invariant version of
Fig.~\ref{fig:boundary-contraction}b) from both left and right; from the
iMPS fixed points, we can then compute physical quantities in the given
state. The value of this tensor network is invariant under gauge transformations $X$,
$X^{-1}$ and $Y$, $Y^{-1}$ applied to the horizontal and vertical indices,
respectively, as shown in Fig.~\ref{fig:gaugefixing}. 
We can now choose the gauges $X$ and $Y$ such as to minimize the amount of
non-positive tensor entries, e.g.\ by maximizing $\big|\sum_{\bm{x}}
A_{\bm{x}}\big|/\sum_{\bm{x}}\big|A_{\bm{x}}\big|$, where the sum runs
over all entries of the double-layer tensor $A_{\bm{x}}$.  

To test the effect of such a gauge fixing in a practically relevant
scenario, we have used a tensor network state (infinite PEPS) which has
been variationally optimized for the ground state of the spin-$\tfrac12$
$J_1$-$J_2$-model on the square lattice for $J_2/J_1=0.5$ (i.e., a
Heisenberg antiferromagnet with coupling $J_1$ along the edges and
coupling $J_2$ along the diagonal).  At this value of $J_2/J_1$, the model
is a highly frustrated topological spin liquid, and exhibits a strong
sign problem in quantum Monte Carlo. For this tensor, we have then studied
how a gauge fixing procedure guided by positivity of the tensor
entries affects the amount of entanglement in the boundary iMPS and thus
the hardness of contraction.
The results are reported in Figure~\ref{fig:gaugefixing} and the figure
caption, where we compare the contraction of the network constructed using
the original double-layer tensor, the tensor where the gauge has been
chosen such as to maximize the hermiticity of the the transfer operator, and the
tensor where the gauge has been chosen such as to maximize the positivity
of the entries while at the same time favoring hermiticity.  We find that
choosing a gauge which maximizes positivity leads to a significant
decrease both of the entanglement entropy and the truncation error of the
boundary iMPS in the contraction.  This demonstrates that the observations
made in our work about the relation of entanglement and positivity also
apply in realistic, translational invariant settings involving systems
with a severe sign problem, and that these insights can thus be used to
improve the accuracy of practical contraction algorithms.

Another way in which the findings reported in this work could be used to 
improve contraction algorithms in practical settings 
builds on the observation made in
Sec.~\ref{sec:pepscontraction-tnr-proof}, where
we showed that in a certain regime, the two-layer tensor
network which appears in PEPS contractions can be transformed into a
positive tensor network by a local TNR-like transformation. In a realistic
regime, we wouldn't expect such a strictly positive TNR decomposition 
to typically exist. However, as we observed, a relatvely small positive bias can
already be sufficient to significantly reduce the complexity in tensor
network contraction. This suggests to construct a TNR-type transformation
whose aim is to give rise to a positively biased tensor network after one
TNR step, Fig.~\ref{fig:cp-contraction}c,d---this could be achieved, e.g., by replacing the separable decomposition in 
Eq.~\eqref{eq:separable-dec} by an approximately separable
decomposition, rather than the singular value decomposition typically used
in TNR. Such a positively biased TNR procedure could then, e.g., be used
as an initial pre-processing step in tensor network contraction, 
thereby lowering the amount of entanglement in the tensor
network, such that the resulting tensor network could subsequently be
contracted much more efficiently.

Lastly, let us give one more argument which illustrates why one should expect findings for
random to be relevant also for tensors which appear in practical
simulations. It is based on the strong concentration effects observed around
the average, an effect which gets very rapidly stronger with bond
dimension~\cite{hayden:generic-ent}.  Thus, a random instance will behave
exactly like the average, with a probability extremely close to one.  If
we now start from a random PEPS and optimize it to be the ground state of
a given Hamiltonian, we will end up with a PEPS in close vicinity of the
true ground state, but which we otherwise (i.e., within this vicinity)
expect to be completely random.  The strong concentration effects thus
imply that with probability rapidly approaching one (for increasing bond
dimension, which is the regime we aim for in variational algorithms), such
a variational PEPS optimization will end up in a PEPS wavefunction which
shares the same properties (in particular regarding the efficient
computability of expectation values) as the random instances studied in
our work. 

Finally, the results of our work strengthen the case of using
entanglement-based algorithms for tensor network contraction, such as
the commonly used
methods based on boundary MPS or the corner transfer matrix, over Monte
Carlo based approaches. 
Firstly, we have observed that in order to reduce the difficulty of
entanglement-based approaches, a significantly lower bias towards positive
entries is required as compared
to Monte Carlo sampling; in fact, the
reduction of difficulty happens earlier for instances with higher bond
dimension (which one would expect to be more difficult, and which have a more severe
sign problem in Monte Carlo). 
In addition,
for Monte Carlo based approaches at any fixed positive bias $\lambda$, the
sign problem will eventually kick in as the system size $N$ is increased,
even though this might happen rather late for larger $\lambda$. In contrast to this, for
entanglement based approaches, the amount of entanglement no longer scales
with $N$ once the positive bias exceeds $\lambda\approx1/D$, and thus,
arbitrarily large systems can be simulated at a cost which only grows
linearly with $N$ or, in the case of infinite systems, becomes system size
independent. Therefore, our findings further strengthen the case to favor
entanglement based methods over Monte Carlo sampling in tensor network
contraction.

\begin{acknowledgments}
We acknowledge helpful conversations with
Garnet Chan,
Ryan Levy,
Daniel Malz,
David Perez-Garcia,
Bram Vanhecke,
and 
Frank Verstraete. We would also like to thank Paul Brehmer  and Anna
Francuz for
providing us with the PEPS tensor for the $J_1$-$J_2$ model.
This research was funded in part by the Austrian Science Fund FWF
(Grant DOIs \href{https://doi.org/10.55776/COE1}{10.55776/COE1}, \href{https://doi.org/10.55776/P36305}{10.55776/P36305}, and \href{https://doi.org/10.55776/F71}{10.55776/F71}), 
the European Union -- NextGenerationEU, and the European Union’s Horizon 2020
research and innovation programme through Grant No.\ 863476 (ERC-CoG SEQUAM). 
\mbox{Jiaqing} Jiang is supported by MURI Grant FA9550-18-1-0161 and the IQIM, an NSF
Physics Frontiers Center (NSF Grant PHY-1125565).
D.H.\ acknowledges financial support from the US DoD through a QuICS Hartree fellowship. 
Jielun Chen is supported by the US National Science Foundation under grant
No.~CHE-2102505.
Part of this work was conducted while the authors were visiting 
the Simons Institute for the Theory of Computing 
(supported by DOE QSA grant \#FP00010905)
during summer 2023 and
spring 2024, whose hospitality we gratefully acknowledge.
The computational results have been in part achieved using the Vienna Scientific Cluster (VSC).
For open access purposes, the authors have applied a CC BY public copyright 
license to any author accepted manuscript version arising from this submission.
\end{acknowledgments}
 
\onecolumngrid

\vspace{1em}
\begin{center}
\rule{12cm}{.2mm}
\end{center}
\vspace{1em}

\twocolumngrid

\begin{appendix}

\section{Detailed discussion of the statmech model obtained from
the Haar-orthogonal model\label{app:statmech-orth}}

In the following, we provide the details of the statmech model for the randomness average of 
the Haar-orthogonal model discussed in Sec.~\ref{sec:effective-model-orth}.

We work with the following $10$-state basis for each site (in this order, read line by line):
\begin{align*}
&(\emptyset)\;;\\
&(12),\;(34),&&(13),\;(24),
&&(14),\ (23)\:;\hspace*{2.5cm}\\
&(12)(34)\:,&&(13)(24)\:,&&(14)(23)\ .
\end{align*}
Note that the three configurations $(12)$, $(34)$, and $(12)(34)$ 
naturally group together, as do the other two triples (typeset on top of each other). 
Also note that the model possesses
a natural $S_4$ symmetry, obtained from permuting the four copies of the
tensor network, which acts by conjugation on the cycles $(ij)$ (and thus,
in particular, keeps those tripes together).

From Eq.~\eqref{eq:4-copy-expansion}
and the expressions for the Haar integrals
(Eq.~\eqref{eq:weingarten-4} and just above),
we immediately obtain the 
weights for the on-site term
\begin{equation}
\tilde V(\sigma) =  (\lambda^4;\ \lambda^2,\dots;\ D^4/(D^4+2),\dots)\ .
\label{eq:appa:on-site}
\end{equation}
For the coupling terms, we need to determine the weight obtained from
overlapping the loop pattern (i.e., $\delta_{x^ix^j}$) corresponding to the
basis elements. Note that a missing pair $(ij)$ in the first 1+7 basis
states amounts to the all-ones tensor, which evaluates to $1$ for every value of $x^i$
and $x^j$. We thus find that for a given edge, any closed loop or open line
contributes a weight $D$ each (cf.\ Fig.~\ref{fig:statmech-mapping}e). 
Carrying out this calculation, we find that the
weight associated to a link with adjacent configurations $\sigma$ and $\sigma'$ is 
\begin{equation}
    \label{eq:appa:edgeweight-unrescaled}
\tilde W(\sigma,\sigma') = \sqrt{Q(\sigma)}W(\sigma,\sigma')\sqrt{Q(\sigma')}\ ,
\end{equation}
where $W(\sigma,\sigma') =
(1/\sqrt{D})^{\tilde k(\sigma,\sigma')}
= e^{-(\log(D)/2)\, \tilde k(\sigma,\sigma')}$,
\begin{equation}
\label{eq:orth-k-def}
\tilde k = \left[\begin{array}{c|cccccc|ccc}
0 &   1 & 1 & 1 & 1 & 1 & 1 &   2 & 2 & 2 \\
\hline
1 &   0 & 2 & 2 & 2 & 2 & 2 &   1 & 3 & 3     \\
1 &   2 & 0 & 2 & 2 & 2 & 2 &   1 & 3 & 3     \\
1 &   2 & 2 & 0 & 2 & 2 & 2 &   3 & 1 & 3     \\
1 &   2 & 2 & 2 & 0 & 2 & 2 &   3 & 1 & 3     \\
1 &   2 & 2 & 2 & 2 & 0 & 2 &   3 & 3 & 1     \\
1 &   2 & 2 & 2 & 2 & 2 & 0 &   3 & 3 & 1     \\
\hline
2 &   1 & 1 & 3 & 3 & 3 & 3 &   0 & 2 & 2     \\
2 &   3 & 3 & 1 & 1 & 3 & 3 &   2 & 0 & 2     \\
2 &   3 & 3 & 3 & 3 & 1 & 1 &   2 & 2 & 0     
\end{array} \right]
\end{equation}
and
$$
Q = (D^{4};\ D^{3},\dots;\ D^{2},\dots)\ .
$$
We can now absorb the weights $\sqrt{Q(\sigma)}$ into the vertex term
$\tilde V(\sigma)$ (there are four $\sqrt{Q(\sigma)}$ adjacent to each vertex) to
obtain a vertex weight 
\[
V(\sigma) = D^4 \times (D^4\lambda^4;\ D^2\lambda^2,\dots;D^4/(D^4+2),\dots)\
.
\]
The prefactor $D^4$ amounts to a rescaling of the partition
function 
and can be compensated by rescaling each tensor $A^{(s)}$ by
$1/D$. After this rescaling, we find that the randomness average of four
copies of the tensor network is equivalent to the partition function of a
statmech model with Hamiltonian
\begin{equation}
\label{eq:app:statmech-totalham}
H({\sigma^{(s)}}) = \sum_s h(\sigma^{(s)}) + \sum_{\langle
s,s'\rangle} k(\sigma^{(s)},\sigma^{(s')})
\end{equation}
at inverse temperature $\beta=1$, where
\[
h(\sigma) = -(4\log(\lambda D);\ 2\log(\lambda D),\dots;
\log\tfrac{D^4}{D^4+2},\dots)\
\]
and 
\[
k(\sigma,\sigma') = \tfrac{\log\,D}{2}\: \tilde k(\sigma,\sigma')\ ,
\]
with $\tilde k$ from Eq.~\eqref{eq:orth-k-def}. Note that $k$ describes a ferromagnetic coupling with additional off-diagonal terms whose Gibbs weights vanish as $D\to\infty$.

\section{Results for the Haar-unitary model: Simulation and effective statmech model}
\label{appendix:unitary-model}

In this Appendix, we discuss the tensor network where the tensors are chosen from a shifted Haar-unitary distribution, that is, 
$U$ in Eq.~\eqref{eq:A-shifted-orthogonal} is chosen from a Haar-random unitary ensemble.

In that case, we can obtain a similar statmech model by replacing the integral on the left hand side of Eq.~\eqref{eq:4-copy-expansion} with the Haar-integral over the unitary group; note that in that case, every second tensor carries a complex conjugate, and thus depends on $\bar U$ rather than $U$. Thus, on the right hand side, only integrals involving an equal number of $U$'s and $\bar U$'s are non-vanishing. Using Weingarten calculus~\cite{collins:weingarten1,collins:weingarten2}, the integral with one $U$ and $\bar U$ evaluates to
$$
\int \mathrm{d}U\, U_{x^1}\otimes \bar{U}_{x^2} = \frac{\delta_{x^1 x^{2}}}{D^4},
$$
and the integral with two pairs evaluates to
\begin{equation}
    \label{eq:weingarten-4-unitary}
\begin{aligned}
    &\int \mathrm{d}U\, U_{x^1}\otimes \bar{U}_{x^2} \otimes U_{x^3}\otimes \bar{U}_{x^4} \\
    =&\frac{1}{D^4(D^4+1)} \left(\delta_{x^1x^2}\delta_{x^3x^4} + \delta_{x^1x^4}\delta_{x^2x^3}\right).
\end{aligned}
\end{equation}
Therefore, for each site, we obtain the 7-state basis
\begin{align*}
&(\emptyset),\\
&(12),(34),&&(14),(23),\hspace*{2.5cm}\\
&(12)(34),&&(14)(23)\ .
\end{align*}
The model differs from the orthogonal group's statmech model only by the number of basis states and the on-site weights of the $(ij)(kl)$ configurations,
\begin{equation}
\tilde V(\sigma) =  (\lambda^4;\ \lambda^2,\dots;\ D^4/(D^4+1),\dots)\ .
\label{eq:appb:on-site}
\end{equation}
Carrying out a similar calculation, we correspondingly obtain 
a coupling weight
\begin{equation}
    \label{eq:appb:edgeweight-unrescaled}
\tilde W(\sigma,\sigma') = \sqrt{Q(\sigma)}W(\sigma,\sigma')\sqrt{Q(\sigma')}\ ,
\end{equation}
with $W(\sigma,\sigma')=e^{-k(\sigma,\sigma')}$,
\begin{equation}
\label{eq:unitary-k-def}
k = \frac{\log\,D}{2}\left[\begin{array}{c|cccc|cc}
0 &   1 & 1 & 1 & 1 &    2 & 2  \\
\hline
1 &   0 & 2 & 2 & 2 &    1 & 3      \\
1 &   2 & 0 & 2 & 2 &    1 & 3      \\
1 &   2 & 2 & 0 & 2 &    3 & 1     \\
1 &   2 & 2 & 2 & 0 &    3 & 1      \\
\hline
2 &   1 & 1 & 3 & 3 &    0 & 2      \\
2 &   3 & 3 & 1 & 1 &    2 & 0        
\end{array} \right] \ ,
\end{equation}
and
\begin{equation}
    Q = (D^4;\:D^3,\dots;\:D^2,\dots)\ .
\end{equation}
After rescaling by $1/D^4$ per site, we obtain the partition function of the Hamiltonian \eqref{eq:app:statmech-totalham} at $\beta=1$, with coupling $k(\sigma,\sigma')$ from Eq.~\eqref{eq:unitary-k-def}, and local field
\[
h(\sigma) = -(4\log(\lambda D);\ 2\log(\lambda D),\dots;
\log\tfrac{D^4}{D^4+1},\dots)\ .
\] 

\begin{figure}
\includegraphics[width=8cm]{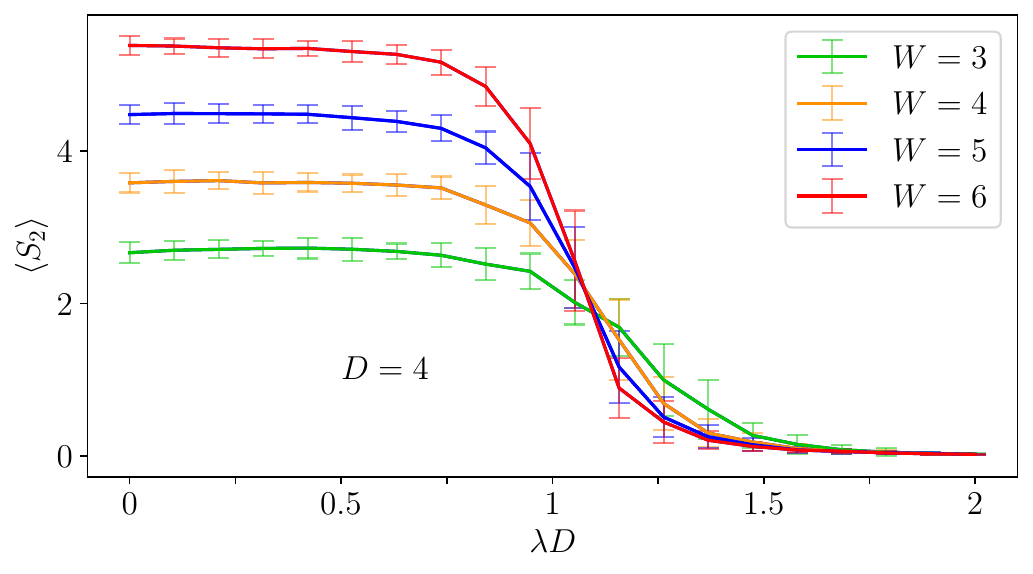}
\caption{
Average $2$-R\'enyi entropy for the Haar-random unitary model, plotted vs.\ $\lambda D$ for $D=4$. Just as for the orthogonal case (Fig.~\ref{fig:ent-scaling}), we find a transition in the entanglement around $\lambda D\approx 1$.
}
\label{fig:unitary_entropy}
\end{figure}

We thus find that the model displays again a ferromagnetic coupling $k$, with off-diagonal terms whose Gibbs weights vanish as $D\to\infty$. In the same limit, the local field 
for $\lambda D<1$ favors the  symmetry broken states $(12)(34)$ and $(14)(23)$, corresponding to the two boundary conditions in Fig.~\ref{fig:statmech-mapping}a, while for $\lambda D>1$, it favors the unique disordered state $(\emptyset)$. This suggests that also the Haar-unitary tensor network should display an entanglement transition around $\lambda D\approx 1$. This is indeed confirmed by numerical simulations for random instances, shown in Fig.~\ref{fig:unitary_entropy}.

\section{The statmech model of rank-one interpolation\label{app:statmech-rank1}
}

In this Appendix, we follow up on the discussion of the interpolation towards a general
(non-positive) rank-$1$ tensor from Sec.~\ref{sec:pos-vs-rank1}, by deriving an effective statmech model for the interpolation towards rank-$1$ tensors which are chosen randomly at each site. We will focus on the Haar-unitary case, but an analogous discussion applies to the Haar-orthogonal scenario.

To this end, recall that in Sec.~\ref{sec:pos-vs-rank1}, we considered interpolations of the form (cf.\ Eq.~\eqref{eq:S-rank1})
\begin{equation}
    \label{eq:appc:A-plus-S}
A^{(s)}_{(x_\ell x_r x_u x_d)} = D^2U^{(s)}_{(x_\ell x_r x_u x_d)}+
\lambda S_{(x_\ell x_r x_u x_d)}^{(s)}\ ,
\end{equation}
where the $S_{(x_\ell x_r x_u x_d)}^{(s)}$ have rank one, 
\begin{equation}
    \label{eq:appc:rank1-S}
S^{(s)}_{x_\ell x_rx_ux_d} = v^\ell_{x_\ell} v^r_{x_r} v^u_{x_u} v^d_{x_d}\ .
\end{equation}
In the following, we will study what happens when all the $v$'s above are chosen independently (for each site $s$ and each direction) at random from the Haar measure. This can be achieved by starting from $S\equiv 1$ 
in~\eqref{eq:appc:A-plus-S}
and multiplying each virtual index of $A^{(s)}$ with a Haar-random unitary; since $U^{(s)}$ itself is chosen Haar-random, it remains Haar-random after that transformation, and we have effectively obtained Haar-random $v$'s in \eqref{eq:appc:rank1-S}. When the indices are contracted, we can replace the pair of Haar-random unitaries on the link by a single Haar-random unitary $u$ (for the same reason, the same also applies to open indices at the boundary). Thus, the resulting model differs from the previous one by a term on each link obtained from averaging over $u\otimes \bar u \otimes u \otimes \bar u$ on the link. 
Weingarten calculus~\cite{collins:weingarten1,collins:weingarten2}
yields the two configurations $\{(12)(34),(14)(23)\}\eqqcolon\mathcal T$ at each side of the link, with corresponding weights
\begin{equation}
Y
=\frac{1}{D^2-1}
\left[\begin{array}{cc}
 1 & -\frac{1}{D}   \\
 -\frac{1}{D} & 1     
\end{array} \right]\ .
\end{equation}
The total weight per link can now be conveniently obtained by observing that the resulting loop patterns on the two sides of $Y$ are a subset of those obtained for the unitary model, namely the last two columns (or rows) of $\tilde W$ in Eq.~\eqref{eq:appb:edgeweight-unrescaled}; let us denote the corresponding $7\times 2$ matrix by $\tilde W_Y$. The partition function of the statmech model is then obtained by contracting the tensor network with $\tilde W_Y Y \tilde W_Y^T$ on the links and a weight $\tilde V$, Eq.~\eqref{eq:appb:on-site}, on the vertices. 

The fact that the link operator is rank $2$ suggests that the model is most compactly parameterized in terms of Ising (i.e., two-valued) variables on the links. 
This can be done in two ways: The first option is to split $Y=XX^T$ with
\[
X = \frac{1}{\sqrt{2(D^2-1)}}
\begin{bmatrix}
1 & 1 \\ 1 & -1 
\end{bmatrix}
\begin{bmatrix}
\sqrt{\frac{D-1}{D}} & 0 \\
0 & \sqrt{\frac{D+1}{D}}
\end{bmatrix}\ ,
\]
which results in a model with Ising variables on the links, coupled by the contraction of $\tilde V$ and four times $\tilde W_Y X$ around each vertex. The resulting model turns out to be an eight-vertex model, though one with a magnetic field, and thus (to our knowledge) not within a class known to be solvable.

A second way to construct a statmech model is to replace $X$ above by the positive square root of $Y$, $Y=R^2$ with $R\ge 0$. The resulting model has Ising variables on the link, and a Gibbs weight $T(i,j,k,l)$ on every vertex (with $i,j,k,l=0,1$ the surrounding link variables) which only depends on the number of $1$'s,
\[
T(i,j,k,l)=w(i+j+k+l)\ .
\]
and where $w(0)=w(4)$, $w(1)=w(3)$. Specifically, after dividing out a factor of $D^4$ per vertex and up to first order in $1/D$, we have
\begin{align}
w(0) = w(4) & = (1+2\lambda^2+\lambda^4)-2\lambda^4\frac1D+\dots\\
w(1) = w(3) & = \lambda^4 - (2\lambda^4-\lambda^2-\tfrac12)\frac1D+\dots\\
w(2) & = \lambda^4 - 2\lambda^4 \frac1D + \dots
\end{align}
In addition, the boundary conditions $(A)$ and $(B)$ which we need to impose to compute the 2-R\'enyi entropy, Fig.~\ref{fig:statmech-mapping}a, are 
obtained by applying $R$ to the two canonical basis vectors, 
and are thus
equal to the $0$ and $1$ Ising configuration up to a $1/D$ correction.

Let us now analyze this model. First, for $\lambda=0$ and $D\to\infty$, the only
non-zero terms are $w(0)=w(4)=1$, that is, the model is an Ising ferromagnet.
For finite but large $D$, we find that in addition, $w(1)=w(3)=1/2D$. The corresponding
tensors can be used to construct a domain wall interfacing the  boundary
conditions $(A)\equiv 0$ and $(B)\equiv 1$, Fig.~\ref{fig:statmech-mapping}a;
this gives rise to a finite cutoff of the entropy, just as seen in
Fig.~\ref{fig:rank1-interpol}a at $\lambda=1$. Since the probability for any of
those interface tensors is $1/2D$, and the length of the domain wall scales
linearly with $W$, we expect a scaling of the entropy proportional to
$W\log D$ for sufficiently large $D$.

Let us next see what happens when we increase $\lambda$, where we will take
$D\to\infty$. Then, we can split $T(i,j,k,l)$ as follows: It is a sum of
(\emph{i}) a ferromagnetic term with weight $1+\lambda^2$, and (\emph{ii}) an
equal weight mixture of \emph{all 16} configurations, that is, a paramagnetic
(disordered) system, with weight $\lambda^4$. We then expect the domain
wall between the $(A)$ and the $(B)$ boundary to break down if the disordered
tensors (suitably normalized) percolate and thus separate the two boundaries.
Naturally, this should happen for a ratio of weights $\tau =
\lambda^4/(1+\lambda^2)$ on the order of $1$; this is indeed consistent with the
observed behavior where the transition in entropy is at a fixed $\lambda$,
independent of $D$.

\end{appendix}

\end{document}